\documentclass[aps,prl,amsfonts,amssymb,twocolumn,amsmath,preprintnumbers,nofootinbib,floatfix,showpacs,superscriptaddress]{revtex4-2}

\usepackage{graphicx}
\usepackage{dsfont}
\usepackage{epstopdf}
\usepackage{color}
\usepackage{hyperref}
\usepackage{bm}
\usepackage{printlen}
\usepackage{units}
\usepackage{bbm,pifont}
\usepackage{tikz}
\usepackage{amsmath}
\usepackage{amssymb}
\usepackage{amsthm}
\usepackage{commath}
\usepackage{amsfonts}
\usepackage[normalem]{ulem}
\usepackage[utf8]{inputenc}
\usepackage{epsfig}
\usepackage{latexsym}
\usepackage{xcolor}
\usepackage{subfigure}
\usepackage{array}
\usepackage{hypcap}
\usepackage{cancel}
\usepackage{braket}
\usepackage{algorithm}
\usepackage{algorithmic}

\newcommand{\Q}{\bm{Q}}
\newcommand{\z}{\hat{\bm{z}}}

\newcommand{\im}{\operatorname{Im}}

\begin{document}

\title{Dissipative Nonlinear Phononics: Nonequilibrium Quasiperiodic Order in Light-Driven Spin-Phonon System}
\date{\today}
\author{Brayan I. Eraso-Solarte}
\author{Yafei Ren}
\email{yfren@udel.edu}
\affiliation{Department of Physics and Astronomy, University of Delaware, Newark, DE 19716, USA}

\begin{abstract}
Nonlinear phononics has emerged as a powerful paradigm for the nonthermal control of quantum materials by engineering a conservative potential energy landscape. Here, we show that dissipation can serve as an additional control knob for nonequilibrium states in nonlinear phononics. We reveal a nontrivial role of dissipation by investigating a spin-phonon coupled system driven by circularly polarized light. By tuning the spin relaxation time $\tau_s$, the steady state undergoes a transition from a trivial limit cycle to a temporally ordered state, which spontaneously breaks the discrete time-translation symmetry imposed by the drive. In this state, both the spin and phonon angular momentum exhibit persistent oscillations at an emergent frequency $\Omega_s$, which is generally incommensurate with the driving frequency. This state is stabilized by a dissipation-induced phase lag between spin and phonon angular momentum that generates a feedback loop sustaining the oscillation. The dissipation-controlled transition can be described within a Landau-type framework using a pseudo-potential, where the order parameter has a $U(1)$ phase symmetry, and its amplitude is proportional to the oscillation amplitude of the phonon angular momentum.
\end{abstract}

\maketitle

Nonlinear phononics has emerged as a powerful paradigm for nonthermal control of quantum materials~\cite{forst2011nonlinear,subedi2021light,de2021colloquium,luo2023large,blank2023two,luo2024terahertz,levchuk2025nonlinear}. Intense terahertz or mid-infrared pulses drive large-amplitude lattice motions that transiently reshape the potential energy landscape~\cite{de2021colloquium}, enabling ultrafast manipulation of magnetic order~\cite{stupakiewicz2021ultrafast,davies2024phononic}, topological band structures~\cite{hubener2018phonon, luo2021light}, and even transient superconducting-like responses~\cite{mitrano2016possible, kennes2017transient, nova2017effective}. These protocols rely on coherent lattice dynamics. While dissipation is intrinsic and unavoidable in such systems, its influence is typically subdominant and detrimental to coherence~\cite{de2021colloquium}. 
In nonlinear phononics, whether dissipation can be leveraged to generate driven orders with no equilibrium analog remains an open question, limiting our understanding of the constructive role dissipation may play.

In this Letter, we demonstrate that dissipation can actively stabilize nonequilibrium order, thereby establishing a novel pathway for engineering nonequilibrium states of matter in nonlinear phononics. We consider a spin-phonon coupled system, which is one of the central platforms for nonlinear phononics and ultrafast spin dynamics~\cite{nova2017effective, Juraschek2022, disa2020polarizing, afanasiev2021ultrafast, stupakiewicz2021ultrafast, mashkovich2021terahertz, luo2023large}. Spin-phonon coupling influences both subsystems: on the phononic side, it gives rise to magnetic-field-tunable phonon spectra~\cite{schaack1975magnetic, thalmeier1977optical} and the phonon Hall effect~\cite{sheng2006theory,zhang2010topological}; on the spin side, optically driven chiral phonons act as effective magnetic fields that modulate magnetization and spin dynamics~\cite{Juraschek2022, luo2023large}. The induced magnetization is an analog of the equilibrium ferromagnetic order, where the coherent phonons modulate the potential energy landscape.

\begin{figure}[h!]
    \centering
    \includegraphics[width=6.3 cm]{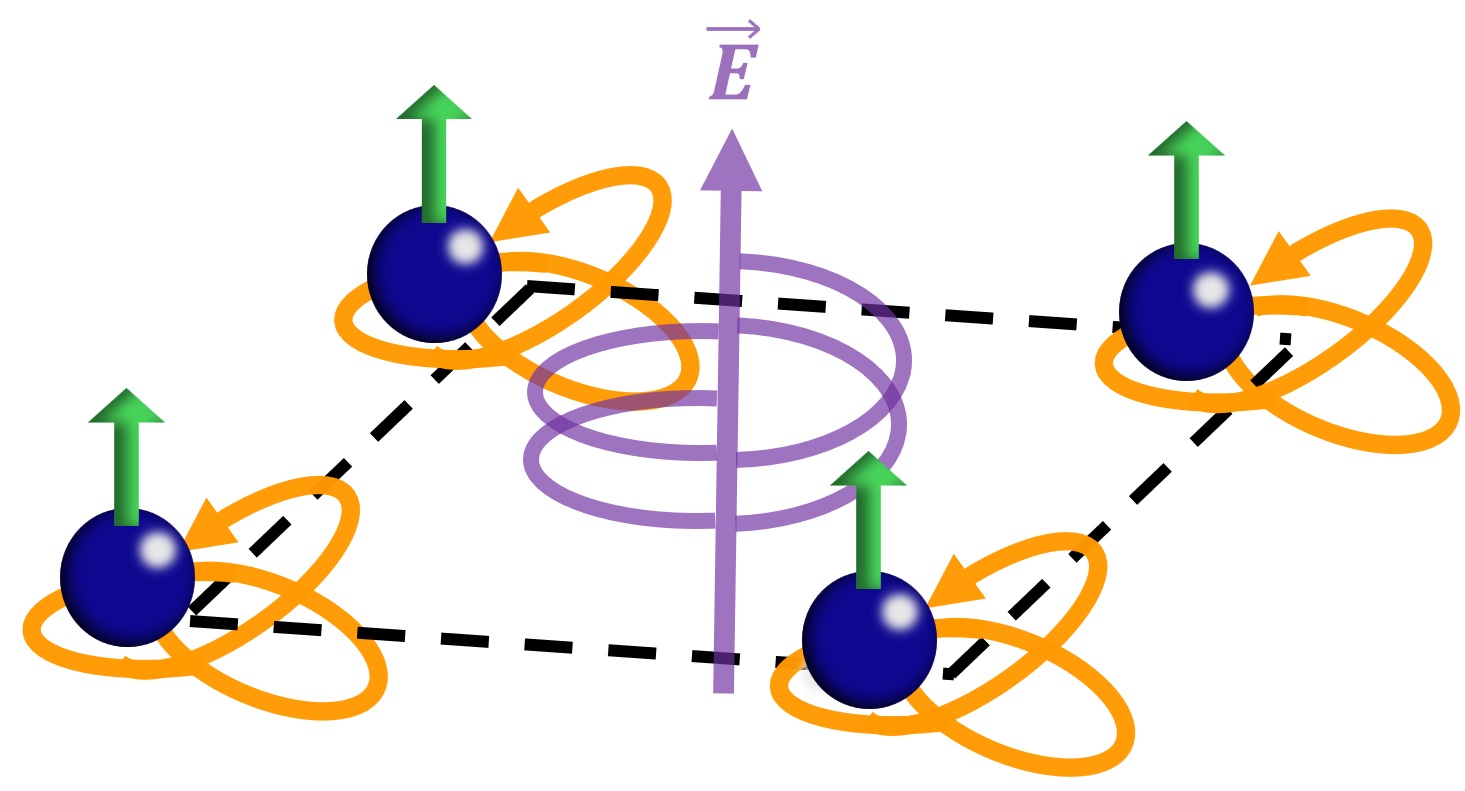}
    \caption{Illustration of the spin-phonon coupled dynamics driven by a circularly polarized light. The steady state shows a longer period than the driving field, breaking the discrete time-translation symmetry. Green arrows indicate local spin.}
   
    \label{Light}
\end{figure}

In contrast, we show that dissipation can modulate the dynamics beyond modifying the potential landscape, stabilizing a robust nonequilibrium temporal order with no equilibrium analogy.
Specifically, we use circularly polarized light to drive coherent chiral phonons with nonzero angular momentum, which subsequently couple to local spins in a paramagnetic phase. 

By controlling the spin relaxation, we find two distinct steady states. A long spin relaxation time stabilizes a limit cycle state, which synchronizes with the driving frequency with a constant spin magnetization. Reducing the spin relaxation time leads to an emerging nonequilibrium state that breaks the discrete time-translation symmetry spontaneously with an oscillating spin magnetization. The oscillation frequency is a fraction of the driving frequency and can be tuned continuously. The oscillating amplitude serves as an effective order parameter, characterizing the transition between the nonequilibrium states. 

We show that the transition is induced by dissipation, which introduces a delayed response between spin and phonon angular momentum, enabling a phase-lag sustained oscillation. 
We then study the effects of driving strength, frequency, and pulsed driving, followed by discussions on material candidates and conditions for experimental observations.

\textit{Spin-phonon dynamics.---} We focus on a generic model system where two degenerate phonon modes, $\bm{Q}=(Q_x, Q_y)$, couple to local spin magnetic moments $\bm{S}$ in a paramagnet through the Raman-type spin-phonon coupling $H_{\rm sp} = \frac{g}{2}\bm{S}\cdot \bm{Q}\times \dot{\bm{Q}}$~\cite{ray1967dynamical,capellmann1989microscopic,sheng2006theory,Juraschek2022,chaudhary2023giant, zhang2014angular, schaack1977magnetic, zhang2010topological}. Here $\bm{Q}\times \dot{\bm{Q}}$ is the phonon angular momentum~\cite{zhang2014angular}. Such coupling has been demonstrated in broad classes of magnetic materials, including both rare-earth magnets and transition-metal compounds,
such as CeF$_3$, CeCl$_3$, and CoTiO$_3$, in both theory and experiment~\cite{schaack1975magnetic, schaack1977magnetic, capellmann1989microscopic, Juraschek2022, luo2023large, chaudhary2023giant, ray1967dynamical, zhang2014angular, sheng2006theory,zhang2010topological, thalmeier1977optical}. This coupling can be strong in the presence of low-lying unoccupied excited states, induced by crystal field splitting in $f$-electron systems~\cite{capellmann1989microscopic} or spin-orbit coupling splitting in $d$-electrons~\cite{chaudhary2023giant}, that can be coupled to the ground-state spin manifold by phonons. 
Here we study infrared-active phonons that can be resonantly excited using light, thanks to recent advancements in terahertz laser~\cite{blank2023two, luo2024terahertz, levchuk2025nonlinear, mciver2020light, stupakiewicz2021ultrafast}. 

We focus on the dynamics driven by a circularly polarized light using the following Lagrangian:
\begin{align}
\label{eq:Lagrangian}
    \mathcal{L}_{\rm ph} = \frac{1}{2} \dot{\bm{Q}}^2 - \frac{\omega^2}{2} \bm{Q}^2 - \frac{g}{2} S\hat{z} \cdot (\bm{Q} \times \dot{\bm{Q}}) + \bm{F}(t) \cdot \bm{Q},
\end{align}
where \(\omega\) denotes the natural oscillation frequency of the degenerate phonon modes, \(g\) represents the spin-phonon coupling constant, and $S$ is the $z$-component of $\bm{S}$. The phonon angular momentum orientation is taken as the $z$-direction that couples to the $z$-component of the local spins. Light couples to the infrared-active phonons via its electric field $\bm{E}$ that generates a force $\bm F = \bm Z \bm E$ with $\bm Z$ the Born effective charge tensor~\cite{Juraschek2022}. We first study a continuous drive with \(\bm{E} = E(\cos (\Omega t),\sin (\Omega t), 0)\) and discuss the effects of laser pulses later. Here \(\Omega = \frac{2\pi}{T_0}\) is the driving frequency with \(T_0\) the period of the light field.

The light-driven phonon motion is obtained by applying the Euler-Lagrange equation to Eq.~\eqref{eq:Lagrangian}, yielding
\begin{align}
\label{EoM}
 \ddot{\bm{Q}} = - \omega^2 \bm{Q} + \bm{F} - g\dot{\bm{Q}}\times  S \hat{z} - \frac{g}{2} \bm{Q}\times \dot{S}\hat{z} - \dot{\bm{Q}}/\tau_p.
\end{align}
Beyond the harmonic restoring force and optically induced driving $\bm{F}=F(\cos\Omega t, \sin\Omega t, 0)$, the phonons acquire a set of spin-mediated effective forces. The term $\dot{\bm{Q}}\times S\hat{z}$ acts as a Lorentz-like force from the local spin, which does not do work. The $\bm{Q}\times \dot{S}\hat{z}$ term captures a dynamical renormalization of the off-diagonal spring constants originating from the time evolution of spin. This term, in contrast, can do work on phonons. Coupling to the environment is incorporated through a relaxation-time approximation, with $\tau_p$ controlling phonon relaxation~\cite{merlin1997generating, dekorsy2006coherent, subedi2014theory, fechner2016effects, juraschek2018sum, Juraschek2022, luo2023large}.

\begin{figure}[t] 
    \includegraphics[width=8.7 cm]{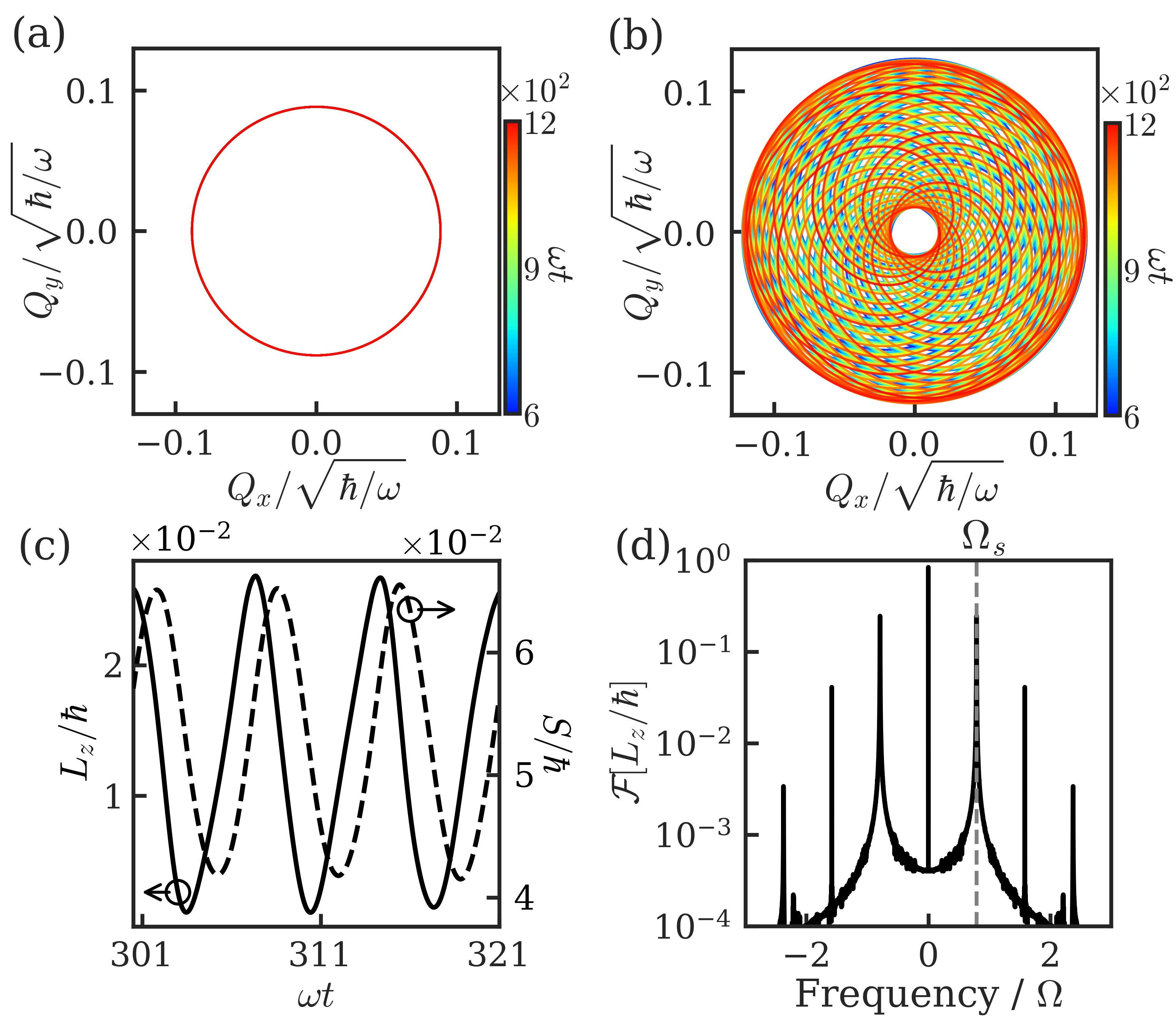}
    \caption{(a) Trajectory of the limit cycle state at \(\omega \tau_s = 32\). (b) Trajectory of the temporally ordered state with \(\omega \tau_s = 4.339\). (c) Oscillating phonon angular momentum $L_z$ and spin $S$. (d) Fourier transform of the angular momentum. \(\Omega_s\) denotes the main oscillation frequency. Other parameters: \(\Omega = 1.16\omega\), \(F = 0.08\omega\sqrt{\hbar \omega}\), \(\chi_p = 4.0 / \hbar\), \(g = \chi_p \left( \frac{10}{1.16^2} \right)\hbar\omega\), \(\omega \tau_p = 10\).} 
    \label{Trajectory_Fourier}
  
\end{figure}

We focus on the paramagnetic regime, where the time evolution of spin is well described by the Bloch equation~\cite{crichton2019practical, bloch1946nuclear, PhysRev.102.104, fletcher1960electron, wangsness1953dynamical}. The $z$-component of the Bloch equation is a relaxational form in which the spin $S(t)$ evolves toward its instantaneous equilibrium value $S_{\rm eq}$ with a phenomenological spin-relaxation time $\tau_s$~\cite{wangsness1953dynamical, stevens1967theory, gurevich2020magnetization, luo2023large}, yielding
\begin{align}
    \label{eq:SpinRelaxation}
    \dot{S}(t) = -\frac{1}{\tau_s}[S(t)-S_{\rm eq}(L_z(t))].
\end{align}
Here, $L_z$ is the phonon angular momentum, and \(S_{\rm eq}(L_z) = \tanh(\chi_p L_z)\) is the thermal expectation value of the spin in the presence of an effective Zeeman field generated by nonzero $L_z$. $\chi_p = g/(2k_B (T-T_c))$ is the paramagnetic susceptibility factor determined by temperature $T$ and the critical temperature $T_c$ for magnetic phase transition. More detailed derivation of the equation is shown in the Supplemental Materials.

\textit{Nonequilibrium Steady State.---} By solving the equations, we identify two distinct nonequilibrium steady states, as shown in Fig.~\ref{Trajectory_Fourier}. Figure~\ref{Trajectory_Fourier}(a) shows that the system relaxes to a trivial steady state in which the phonons form a limit cycle following the circularly polarized drive. The phonon period is the same as the driving period with a constant angular momentum. The corresponding spin is also a constant value determined by the angular momentum. 

\begin{figure}[t] 
\centering \includegraphics[width=8.7cm]{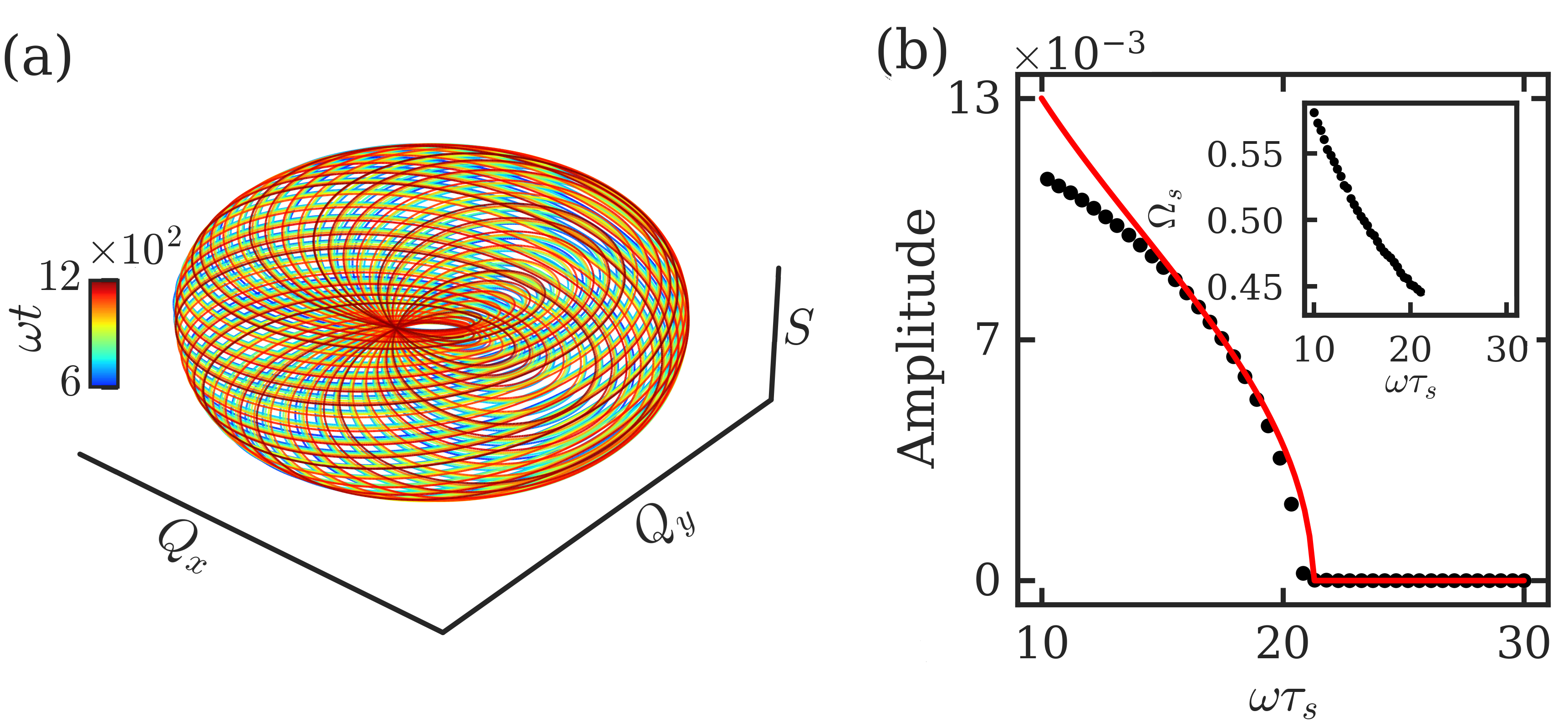} \caption{(a) Three-dimensional trajectory of $(Q_x,Q_y,S)$ that winds around a torus with parameters the same as those in Fig.~\ref{Trajectory_Fourier}(b). (b) Amplitude of $\mathcal{F}(L_z)$ at $\Omega_s$ (see the inset) as a function of $\tau_s$. Black dots show numerical results, and the red line shows analytical results. Here $\chi_p = 4/\hbar$, $g=\chi_p(10/1.16^2)\hbar\omega$, $\Omega = 1.16\omega$, $F = 0.08\omega\sqrt{\hbar \omega}$, and $\omega\tau_p = 10$.}
\label{fig:3d} 
\end{figure}

As the spin relaxation time decreases, the steady state undergoes a qualitative transition. As shown in Fig.~\ref{Trajectory_Fourier}(b), the phonon trajectory becomes quasiperiodic, with a characteristic period longer than that of the external drive. The resulting steady state, therefore, breaks the discrete time-translation symmetry imposed by the driving field. In this regime, both the phonon angular momentum $L_z$ and the spin $S$ exhibit persistent periodic oscillations, as shown in Fig.~\ref{Trajectory_Fourier}(c).
The Fourier spectrum of $L_z$, displayed in Fig.~\ref{Trajectory_Fourier}(d), contains a series of discrete peaks. We denote the dominant nonzero frequency as $\Omega_s$, while the remaining peaks appear at its integer multiples $n\Omega_s$. Importantly, $\Omega_s$ is smaller than the driving frequency $\Omega$ and is not constrained to be commensurate with it, confirming the quasiperiodic nature of the steady state.
The nature of the trajectory becomes clearer in the three-dimensional $(Q_x,Q_y,S)$ phase space, as illustrated in Fig.~\ref{fig:3d}(a). In this representation, the trajectory densely winds around a torus, whose projection onto the $Q_x$-$Q_y$ plane corresponds to the trajectory shown in Fig.~\ref{Trajectory_Fourier}(b).

\textit{Dissipation-Induced Feedback and Instability.---}
The emergent temporal order is stabilized by the phase delay between $L_z$ and $S$ induced by dissipation. To understand this explicitly, we analyze the balance of the energy dissipation and energy input in the steady state. We first consider the limit-cycle state where $\bm Q$ oscillates at the same frequency as the driving force. The energy dissipation from $-\dot{\bm Q}/\tau_p$ is balanced by the work done by $\bm F$, since the restoring force is conservative, the effective Lorentz force does not do work, and the force from the $\dot S$ term vanishes.
In the temporally ordered state, $\bm Q$ has additional oscillations at frequencies of $\Omega + n \Omega_s$. The additional energy dissipation should also be balanced by the work done by the forces. 

This work cannot be supplied by the external driving, as it is off-resonant with the temporal oscillation. The only available source is the feedback force $-\frac{g}{2}\bm{Q}\times \dot{S}\hat z$. The corresponding power is
$P=-\frac{g}{2}\bm{Q}\times \dot{S}\hat z\cdot \dot{\bm Q}
= \frac{g}{2}\dot{S}L_z$.
If $S$ and $L_z$ oscillate in phase at the same frequency, the corresponding work done over one period vanishes. However, when a finite phase difference develops between them, as illustrated in Fig.~\ref{Trajectory_Fourier}(c), the net work over one cycle becomes nonzero. This finite work offsets the additional dissipation, thereby enabling an instability toward a temporally ordered state.

The phase difference between $S$ and $L_z$ is controlled by the spin relaxation time $\tau_s$ as shown below. Assuming $S_{\rm eq}\simeq \chi_p L_z$, the spin dynamics governed by Eq.~\eqref{eq:SpinRelaxation} yields
\begin{align}
S(t)=
\chi_p\int_{-\infty}^{t}\frac{d t'}{\tau_s}
e^{-(t-t')/\tau_s} L_z(t') .
\end{align}
The retarded kernel $e^{-(t-t')/\tau_s}$ introduces a phase lag between $S$ and $L_z$. Writing the oscillatory components of $L_z$ and $S$ as the real parts of $\delta L_z e^{i\Omega_s t}$ and $\delta S e^{i\Omega_s t}$, respectively,
one finds
$\delta S=\frac{\chi_p}{1+i\Omega_s\tau_s}\,\delta L_z$. The dynamical spin-phonon susceptibility $\frac{\chi_p}{1+i\Omega_s\tau_s}$ determines the phase lag, $\arg(1+i\Omega_s\tau_s)$. The net work performed over one period is proportional to
$\frac{\Omega_s^2 \tau_s}{1+(\Omega_s\tau_s)^2}$~\cite{SM}.
In the large-$\tau_s$ limit, the net work vanishes due to the suppression from the denominator.

\textit{Dynamical Transition.---} 
To further understand the effect of the dissipation, we characterize the $\tau_s$ induced dynamical transition between the two states by computing the Fourier spectrum $\mathcal{F}(L_z)$ of the phonon angular momentum. 

The amplitude of $\mathcal{F}(L_z)$ at the emergent frequency $\Omega_s$ is plotted by the dots in Fig.~\ref{fig:3d}(b). For large $\tau_s$, the amplitude remains vanishingly small, indicating a limit-cycle steady state. As $\tau_s$ reduces and crosses a critical value $\tau_{s}^c$, the amplitude increases sharply, signaling the onset of the emergent temporal order. In addition to the amplitude, the oscillation frequency $\Omega_s$ also varies continuously as a function of $\tau_s$, as shown in the inset of Fig.~\ref{fig:3d}(b). 
Therefore, dissipation can be used as a control parameter to actively induce and stabilize emergent nonequilibrium orders. 

To better understand the transition, we introduce a complex variable $\psi=Q_x + iQ_y$ with a trial solution $\psi = (A+B(t))e^{i\Omega t}$ where $A$ does not change over time. 
At $B=0$, the solution corresponds to the limit cycle state with constant angular momentum and spin, where $A$ can be solved self-consistently~\cite{SM}. A nonzero $B(t)$ indicates the emergence of the temporal order.
We set $B(t) \simeq B_0e^{i\Omega_s t}$ by neglecting higher harmonics of $\Omega_s$ that are relatively weak as shown in Fig.~\ref{Trajectory_Fourier}(d). 
A nonzero $B_0$ leads to an oscillating phonon angular momentum $\delta L_z e^{i\Omega_s t} $ with $\delta L_z= A^*B_0(2\Omega + \Omega_s)$ and oscillating spin $\frac{\chi_p}{1+ i \Omega_s \tau_s} \delta L_z e^{i\Omega_s t}$ with a phase lag.

Substituting the trial solution into the Eq.~\eqref{EoM}, we find the amplitude equation for $B(t)$:
\begin{align}\label{eq:amplitude}
    \ddot{B} +\Gamma_0 \dot{B} + (\Omega_0^2 - \mathcal{S}_0) B +\Gamma^{\prime}|B|^2 \dot{B} + \Omega^{\prime\: 2}|B|^2 B =0
\end{align}

where $\Gamma_0= \frac{1}{\tau_p}+i2\Omega
-i g\chi_p\Omega|A|^2$, $\Omega_0^2= \omega^2+\Omega^2 + i\Omega \Gamma_0$,  $\Gamma^{\prime}=-i g\chi_p(\Omega+\Omega_s)$,  $\Omega^{\prime\: 2}= i\Omega \Gamma'$, and $\mathcal{S}_0=\frac{-g\chi_p}{1+i\Omega_s\tau_s}
\left(\frac{2\Omega+\Omega_s}{2}\right)^2 |A|^2$.

This amplitude equation always has a trivial solution $B=0$ for the limit cycle state. 

To obtain nontrivial solutions for the temporal order with $B\neq 0$, we substitute $ B(t) \simeq B_0e^{i\Omega_s t}$ into Eq.~\eqref{eq:amplitude} to obtain a complex equation for $|B_0|^2$ and $\Omega_s$. The imaginary part of the equation determines $\Omega_s$. By substituting $\Omega_s$ into the real part of the equation, we find the solution 
\begin{align} \label{eq:B0f}
    |B_0|^2=f(\tau_s),
\end{align} 
with $f(\tau_s)$ a complicated expression depending on $\tau_s$ and other parameters as detailed in the Supplemental Materials~\cite{SM}. Only positive $f(\tau_s)$ gives rise to physically meaningful solutions. At large $\tau_s$, we find that $f(\tau_s)$ is negative. As $\tau_s$ reduces, $f(\tau_s)$ crosses zero linearly at the critical value $\tau_s^c$. Thus, near the transition, the amplitude increases with a square root shape: $ |B_0|\propto \sqrt{\tau_s^c-\tau_s }$. The oscillating angular momentum obtained from the analytical solution of $B_0$ agrees quantitatively with the numerical results near the transition point. Deep into the temporally ordered regime for smaller $\tau_s$, the analytical solutions deviate from the numerical solutions as the higher harmonic components of $B(t)$ start to play a role.

\begin{figure}[t] 
\centering \includegraphics[width=8.7cm]{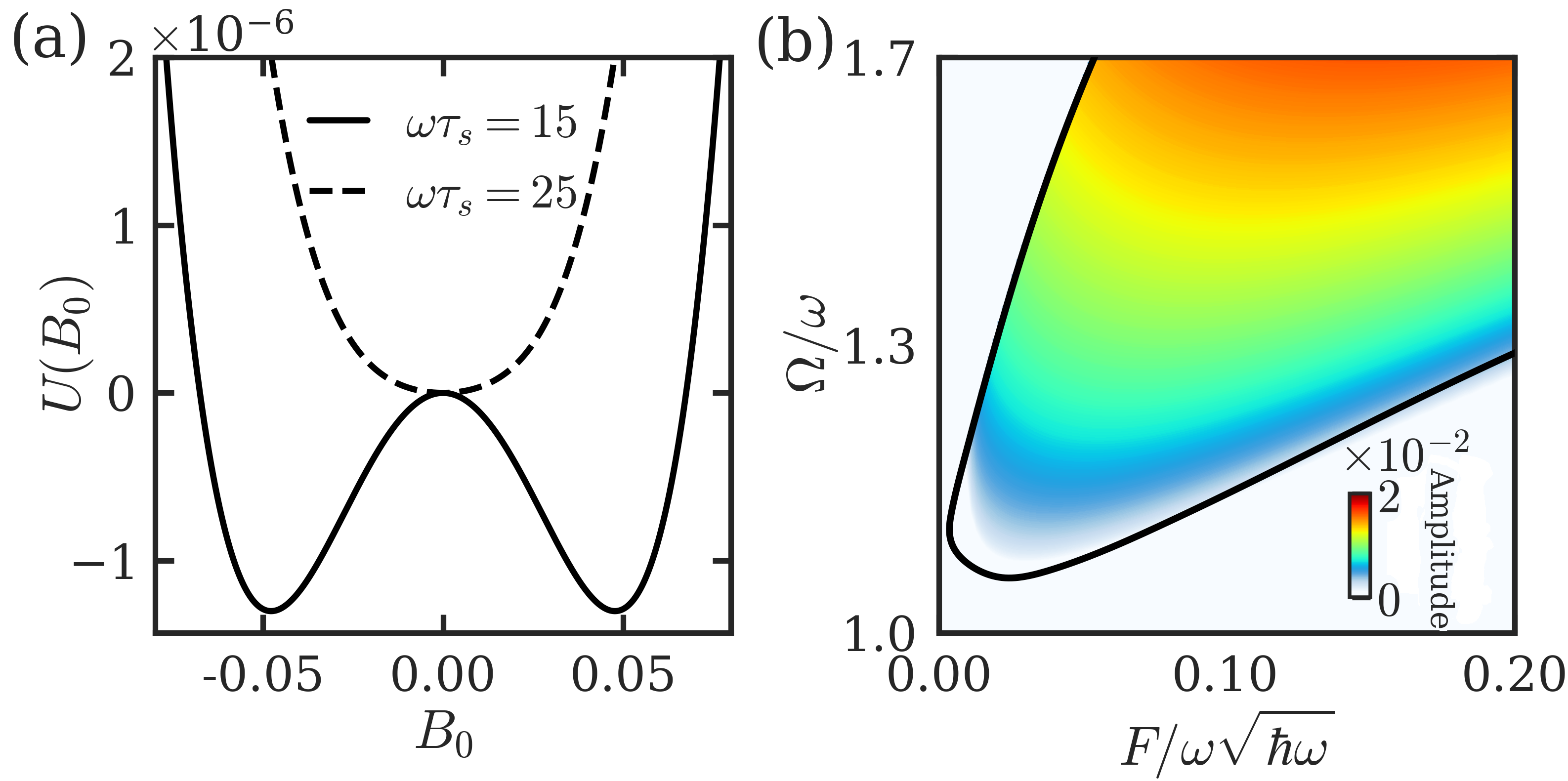} \caption{(a) Pseudo-potential for the effective Landau phase transition theory as a function of the ``order parameter'' $B_0$ at different $\tau_s$. (b) Dynamical phase diagram vs driving frequency \(\Omega\) and driving force \(F\). Color map shows the amplitude of the $\mathcal{F}(L_z)$ at $\Omega_s$. The solid black line indicates the analytically predicted phase boundary. Here $\chi_p = 4.0/\hbar$, $g=\chi_p(10/1.16^2)\hbar\omega$, and $\omega\tau_p = 10$. Panel (a) uses $\Omega = 1.16\omega$ and $F = 0.08\omega\sqrt{\hbar \omega}$; and panel (b) uses $\omega \tau_s=20$.}
\label{Eigenvalues} 
\end{figure}

\textit{Nonequilibrium Landau Theory of Temporal Order.---} The dynamical transition is reminiscent of the Landau phase transition with an order parameter $|B_0|$. The trivial solution of $B_0=0$ and nontrivial solutions in Eq.~\eqref{eq:B0f} can be captured by a Landau-like pseudo-potential 
\begin{align}
    U(|B_0|) = \frac{1}{4}|B_0|^4 - \frac{1}{2} f |B_0|^2
\end{align}
such that $ \frac{\partial  U}{\partial |B_0|}= |B_0|(|B_0|^2 -f) =0$ determines the possible steady states. In Fig.~\ref{Eigenvalues}(a), we plot the potential at two different $\tau_s$. At the larger $\tau_s$, the potential has one local minimum at $B_0=0$, indicating a limit cycle state. At the smaller $\tau_s$, the potential becomes a Mexican-hat shaped potential with local minima at nonzero $|B_0|$, indicating a temporally ordered state. The state possesses a continuous $U(1)$ symmetry associated with the phase of the complex amplitude $B_0$, which reflects the freedom to choose the phase of the emergent oscillation.

The pseudo-potential also provides a practical criterion to locate the transition controlled by other parameters that $f$ depends on.
Figure~\ref{Eigenvalues}(b) shows the dynamical phase diagram in the $(F,\Omega)$ parameter space.  
Using the pseudo-potential $U(B_0)$, we can find the phase boundary at $f(F,\Omega)=0$. 
The resulting analytical boundary is shown by the solid black line in Fig.~\ref{Eigenvalues}(b), which agrees with the numerical phase diagram across the studied parameter space. The color map in this figure shows the amplitude of the angular momentum oscillation in the temporally ordered state.

\begin{figure}[t] 
\centering \includegraphics[width=8.7cm]{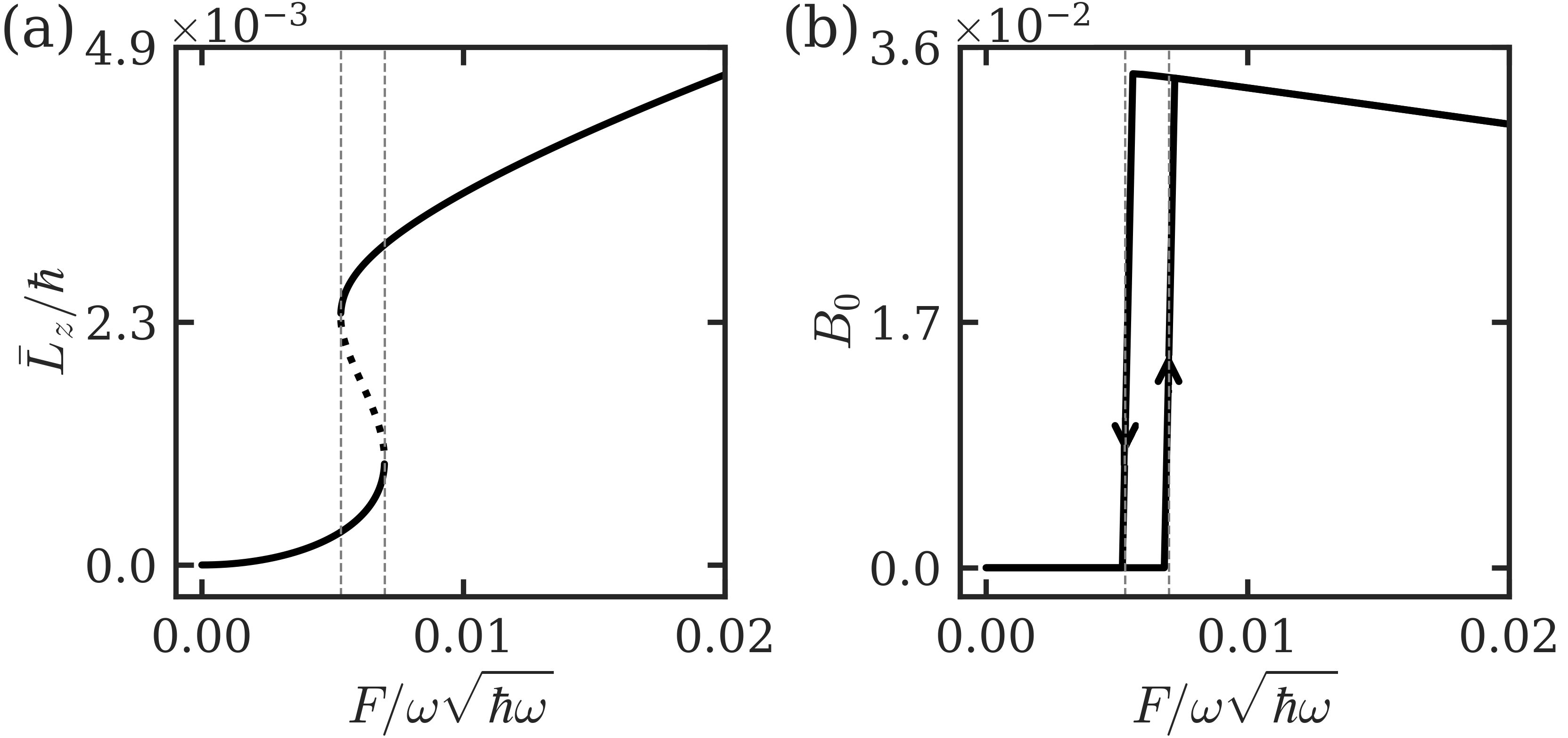}  \caption{(a) Constant part of the phonon angular momentum, $\bar{L}_z$, as a function of $F$. Solid lines show stable solutions, and the dotted line shows unstable solutions. The realization of the stable solution depends on the initial condition, leading to (b) the hysteresis behavior of $B_0$ by sweeping $F$ (arrows indicate the sweep direction). Here $g=\chi_p\!\left(\frac{10}{1.16^2}\right)\hbar\omega$, $\chi_p=4.0/\hbar$, $\omega\tau_s=20$, $\omega\tau_p=10$, and $\Omega=1.16\,\omega$.}
\label{Phase_F_O} 
\end{figure}

\textit{Dynamical Hysteresis.---} The dynamical transition can also form hysteresis by sweeping the driving force. To see so, we first self-consistently solve $A$ in the trial solution $\psi$, which gives rise to the constant part of the angular momentum $\bar{L}_z$~\cite{SM}. Figure~\ref{Phase_F_O}(a) plots the solutions of $\bar{L}_z$ as a function of the driving force $F$. We find that at weak and strong driving forces, there is only one solution in $\bar{L}_z$, indicating one stable solution in the steady state. In the intermediate region between the dashed gray lines, however, there are three solutions, indicating that there are two stable solutions as shown in solid lines and one unstable solution as shown in a dotted line. The smaller $\bar{L}_z$ gives rise to a zero solution of $B_0$ as shown in Fig.~\ref{Phase_F_O}(b), whereas the larger $\bar{L}_z$ gives rise to a nonzero $B_0$. As a result, if one can continuously sweep the driving strength $F$, a hysteresis loop can be observed.

\begin{figure}[t] 
\centering \includegraphics[width=8.7cm]{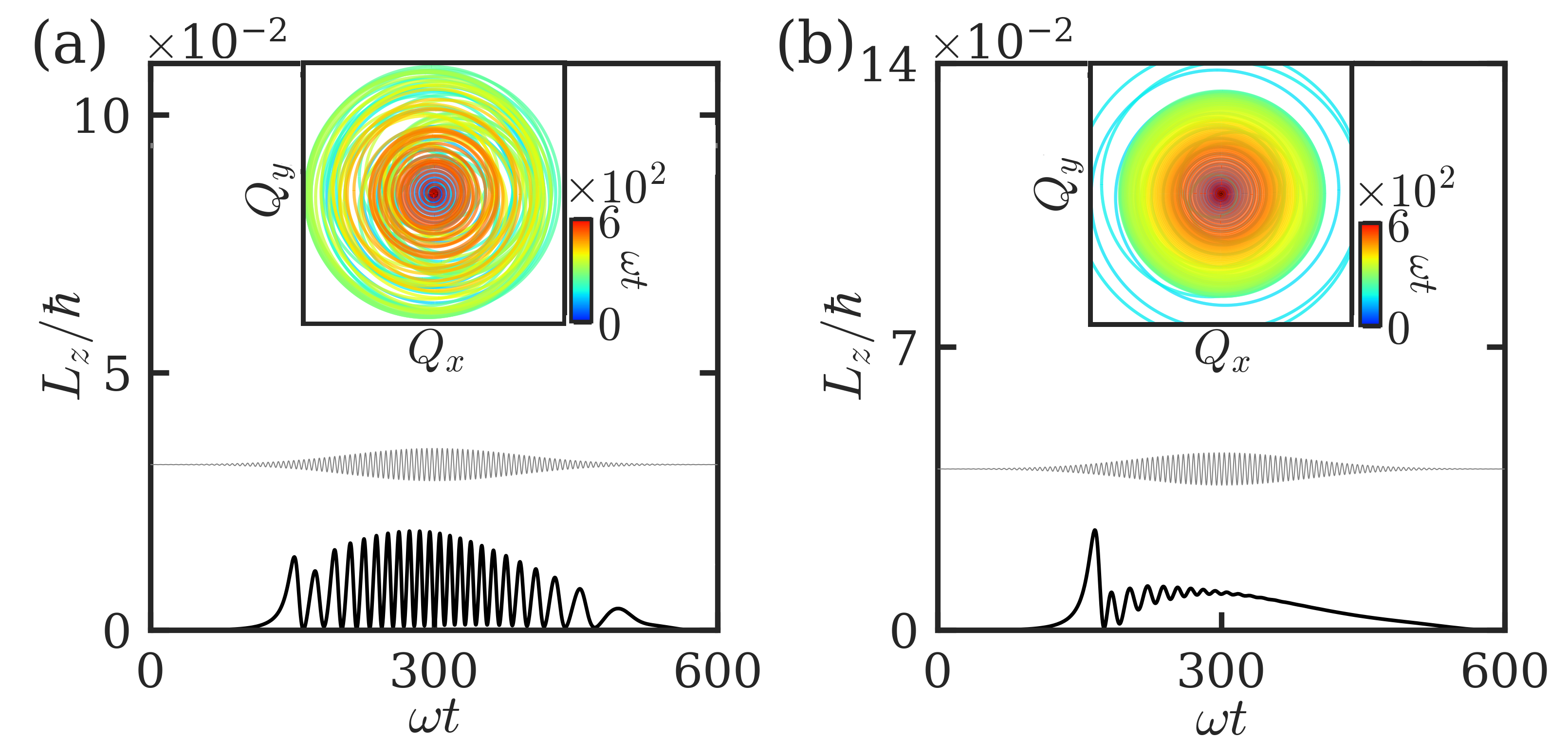}  \caption{Trajectory and angular momentum under pulsed driving at (a) \(\omega \tau_s = 15\) and (b) \(\omega \tau_s = 40\). Here \(\Omega = 1.16\omega\), \(F = 0.08\omega\sqrt{\hbar \omega}\), \(\chi_p = 4.0 / \hbar\), \(g = \chi_p \left( \frac{10}{1.16^2} \right)\hbar\omega\), \(\omega \tau_p = 10\), \(\omega \tau_{\text{pulse}} = 210\), and $\omega t_0=300$. {The grey curve schematically shows the $x$-component of the laser pulse.} 
} \label{pulse} 
\end{figure}

\textit{Pulsed Driving Forces.---}
The temporally ordered state can also emerge under pulsed driving, which is more relevant for realistic experimental conditions~\cite{Juraschek2022, luo2023large}. A Gaussian laser pulse generates a pulsed driving force of the form
\begin{equation}
\label{eq:pulse}
\resizebox{7.5cm}{!}{%
$\displaystyle
\bm{F}(t) =
F \left(
\cos \Omega t , \,
\sin \Omega t , \,
0
\right) 
\exp\left(-\frac{(t-t_0)^2}{2\left(\tau_{\text{pulse}}/\sqrt{8\ln(2)}\right)^2}\right)$
}
\end{equation}
where \(F\) represents the maximum amplitude of the pulse, \(\Omega\) is its central frequency, \(\tau_{\text{pulse}}\) defines the pulse duration through its full-width at half maximum, and $t_0$ sets the time of the pulse peak. 
Depending on the spin relaxation time $\tau_s$, we observe two distinct dynamical responses. For a shorter relaxation time, the system develops a temporally ordered state, as shown in Fig.~\ref{pulse}(a). In this regime, the phonon angular momentum exhibits persistent oscillations throughout the entire time evolution. The oscillation period is longer than the period of the driving force, indicating the emergence of the temporal order. In contrast, for longer spin relaxation times, the behavior changes qualitatively. As shown in Fig.~\ref{pulse}(b), the angular momentum initially oscillates, but the oscillation amplitude decays with time. This transient oscillation is consistent with interference between the driving frequency and the intrinsic phonon dynamics during the establishment of the steady state. After the transient regime, the angular momentum becomes smooth and non-oscillatory, indicating that the system relaxes to a limit-cycle state synchronized with the driving field.

\textit{Summary.---} We demonstrated that dissipation can be employed to actively control the nonequilibrium steady state in nonlinear phononics by studying a spin-phonon coupled system driven by circularly polarized light. Two distinct steady states are established by varying the spin relaxation time. One is a limit cycle state with constant phonon and spin angular momentum, while the other is a temporally ordered state where the phonon and spin angular momentum oscillate at an emergent frequency $\Omega_s$ that is a fraction of the driving frequency. The latter is stabilized by the dissipation-induced phase lag between the phonon angular momentum and the spin subsystem that forms a feedback loop. The dynamical transition between them is well described by an amplitude equation for the phonon amplitude $B_0$ at the emergent frequency. The transition is effectively described by a Landau-type pseudo-potential in terms of the order parameter $B_0$, which is a complex order parameter with a $U(1)$ phase-rotation symmetry.

The predicted phenomena are potentially observable in experiments. Strong spin-phonon coupling has been reported in a variety of materials, including f-electron systems such as $\rm CeCl_3, ~CeF_3, ~PrCl_3, ~and ~NdCl_3$~\cite{schaack1976observation, schaack1977magnetic, schaack1977magnetic1, thalmeier1977optical, ahrens1979phonon, capellmann1989microscopic, strohm2005phenomenological, 4f1983LiTbF4}, as well as $d$-electron systems such as CoTiO$_3$\cite{chaudhary2023giant}. The required driving strength is experimentally feasible: the threshold driving force $F$ is estimated to be below $0.01\,\omega\sqrt{\hbar\omega}$, which lies within the range achievable in recent pump-probe experiments~\cite{Juraschek2022, luo2023large, ren2024light, Afanasiev2021}. 
Regarding spin relaxation, our analysis indicates that the temporally ordered state requires a spin relaxation time that is several times longer than the phonon lifetime. Recent experiments report a spin relaxation approximately 50 times longer than that of phonons at low temperatures~\cite{luo2023large}, suggesting that the required regime is experimentally accessible. The spin relaxation time can be controlled through several routes. Introducing magnetic impurities or doped carriers provides an efficient way to shorten the spin lifetime~\cite{jiang2009electron}. Alternatively, materials with higher magnetic transition temperatures allow the paramagnetic phase to be studied at elevated temperatures, where spin relaxation time is naturally reduced~\cite{luo2023large, lujan2024spin}. A complementary strategy is to reduce the phonon frequency, which increases the phonon lifetime and effectively enlarges the ratio between the spin and phonon relaxation times.

\begin{acknowledgments}
\textit{Acknowledgements.---}This research was supported by the US National Science Foundation (NSF) Grant No. 2531815.
\end{acknowledgments}

\clearpage
\thispagestyle{empty}
\null
\newpage

\onecolumngrid

\setcounter{section}{0}
\setcounter{subsection}{0}
\setcounter{equation}{0}
\setcounter{figure}{0}
\setcounter{table}{0}

\renewcommand{\theHequation}{SM.\arabic{equation}}
\renewcommand{\theHfigure}{SM.\arabic{figure}}
\renewcommand{\theHtable}{SM.\arabic{table}}
\renewcommand{\theHsection}{SM.\arabic{section}}

\begin{center}
{\large\bfseries Supplemental Material}\\[0.5em]
{\bfseries for}\\[0.3em]
{\large\itshape Dissipative Nonlinear Phononics: Nonequilibrium Quasiperiodic Order in Light-Driven Spin-Phonon System}\\[1em]
Brayan I. Eraso-Solarte and Yafei Ren\\
Department of Physics and Astronomy, University of Delaware, Newark, DE 19716, USA
\end{center}

\section{Deriving the longitudinal Bloch equation from a Lindblad master equation}
\label{subsec:SM_lindblad_to_bloch}

To connect the spin-relaxation dynamics to a microscopic model, we derive the longitudinal Bloch equation used in the main text starting from a Markovian Lindblad master equation for an effective two-level system. We consider a two-level system with instantaneous Zeeman splitting $\omega_s(t)$ along $\hat{\bm z}$,
\begin{equation}
H(t)=-\frac{\omega_s(t)}{2}\,\sigma_z,
\end{equation}
and define the longitudinal polarization
\begin{equation}
S(t)\equiv \langle \sigma_z\rangle=\mathrm{Tr}\!\left[\rho(t)\sigma_z\right].
\end{equation}
The reduced dynamics in the weak-coupling, Markovian limit is modeled by
\begin{equation}
\dot\rho = -i[H,\rho]+\Gamma_{\downarrow}(t)\,\mathcal D[\sigma_-]\rho+\Gamma_{\uparrow}(t)\,\mathcal D[\sigma_+]\rho,
\qquad
\mathcal D[A]\rho \equiv A\rho A^\dagger-\frac12\{A^\dagger A,\rho\},
\end{equation}
where $\Gamma_{\downarrow}$ and $\Gamma_{\uparrow}$ are the downward (relaxation) and upward (excitation) rates, respectively.

Since $H(t)$ is diagonal in the $\sigma_z$ eigenbasis, the Hamiltonian commutator does not change the populations; therefore it does not contribute to $\dot S(t)$. The evolution of $S(t)$ is entirely determined by the dissipators.

Evaluating the population dynamics generated by the two dissipators and using $\rho_{\uparrow\uparrow}=(1+S)/2$ and $\rho_{\downarrow\downarrow}=(1-S)/2$, one obtains a closed first-order equation,
\begin{equation}
\dot S(t)=-(\Gamma_{\uparrow}+\Gamma_{\downarrow})\,S(t)+(\Gamma_{\uparrow}-\Gamma_{\downarrow}).
\end{equation}
This can be rewritten in the standard longitudinal Bloch form
\begin{equation}
\label{eq:SM_bloch_from_lindblad}
\dot S(t)=-\frac{1}{\tau_s}\Big[S(t)-S_{\rm eq}(t)\Big],
\end{equation}
by identifying
\begin{equation}
\label{eq:SM_tau_and_Seq}
\frac{1}{\tau_s}\equiv \Gamma_{\uparrow}(t)+\Gamma_{\downarrow}(t),
\qquad
S_{\rm eq}(t)\equiv \frac{\Gamma_{\uparrow}(t)-\Gamma_{\downarrow}(t)}{\Gamma_{\uparrow}(t)+\Gamma_{\downarrow}(t)}.
\end{equation}

If the environment is at temperature $T$, detailed balance relates the rates through
\begin{equation}
\frac{\Gamma_{\uparrow}(t)}{\Gamma_{\downarrow}(t)}=e^{-\beta \omega_s(t)},
\qquad \beta=\frac{1}{k_B T}.
\end{equation}
Inserting this relation into Eq.~\eqref{eq:SM_tau_and_Seq} yields
\begin{equation}
\label{eq:spin_equilium}
S_{\rm eq}(t)=\tanh\!\left(\frac{\beta\,\omega_s(t)}{2}\right),
\end{equation}

In our problem the spin experiences an effective Zeeman field generated by the phonon angular momentum,
\begin{equation}
H_Z(t)=-B_{\rm eff}(t)\,S \equiv -g\,L_z(t)\,S,
\qquad\Rightarrow\qquad
\omega_s(t)=B_{\rm eff}(t)=g\,L_z(t),
\end{equation}
so that the equilibrium polarization becomes
\begin{equation}
\label{eq:SM_seq_lz}
S_{\rm eq}\big(L_z(t)\big)=\tanh\!\left(\frac{g\,L_z(t)}{2k_B T}\right)\equiv \tanh\!\big(\chi_p\,L_z(t)\big),
\qquad
\chi_p=\frac{g}{2k_B T}.
\end{equation}
Near the stability boundary, where $|\chi_p L_z|\ll 1$, we use the linear approximation
\begin{equation}
S_{\rm eq}(L_z)\simeq \chi_p\,L_z,
\end{equation}
while keeping $\tau_s$ as the phenomenological longitudinal relaxation time set by the bath through $1/\tau_s=\Gamma_{\uparrow}+\Gamma_{\downarrow}$.

\section{Floquet stability of the constant-angular-momentum solution}

\label{sec:supp_floquet_stability}

\subsection{Trial solution under circularly polarized driving}
\label{subsec:trial_solution_cp}

We write the phonon displacement as the real part of a complex amplitude $\mathbf{Q}$:
\begin{align}
\label{eq:trial_solution_cp}
(Q_x,Q_y)
&=
\text{Re}\left[
\bar{Q}\,(1,-i)\
e^{i\Omega t}
\right],
\end{align}
where $\bar{Q}=|\bar{Q}| e^{i\theta}$ is a complex constant amplitude.
The relative phase $-i=e^{-i\pi/2}$ encodes a $90^\circ$ phase lag between the $x$ and $y$ components, corresponding to circular motion in the $xy$ plane.
In the synchronized state, the dynamics is strictly periodic with the driving period
\begin{equation}
T_0=\frac{2\pi}{\Omega}.
\end{equation}

Taking time derivatives of the ansatz yields
\begin{equation}
\dot{\mathbf{Q}} = i\Omega 
\bar{Q}\,(1,-i)\
e^{i\Omega t},
\qquad
\ddot{\mathbf{Q}} = -\Omega^2 
\bar{Q}\,(1,-i)\
e^{i\Omega t}.
\end{equation}

The equation of motion for the phonon coordinate is
\begin{equation}
\ddot{\mathbf{Q}}
=
-\omega^2 \mathbf{Q}
+\mathbf{F}
- g\,\dot{\mathbf{Q}}\times \bar{S}\hat{z}
- \frac{g}{2}\mathbf{Q}\times \dot{\bar{S}}\hat{z}
- \frac{\dot{\mathbf{Q}}}{\tau_p}.
\end{equation}
In the steady state of the synchronized state the spin polarization is time independent, $\dot{S}=0$, and the fourth term vanishes.
Solving for the driving force, we obtain
\begin{equation}
\mathbf{F}
=
\ddot{\mathbf{Q}}+\omega^2\mathbf{Q}
+ g\,\dot{\mathbf{Q}}\times \bar{S}\hat{z}
+ \frac{\dot{\mathbf{Q}}}{\tau_p}.
\end{equation}

Evaluating the cross product explicitly,
\begin{equation}
g\,\dot{\mathbf{Q}}\times \bar{S}\hat{z}
=
i g \bar{S}\Omega 
\bar{Q}\,(-i,-1)\
e^{i\Omega t}
=
- g \bar{S}\Omega\,\sigma_y\,\mathbf{Q},
\end{equation}
and collecting all terms, the force can be written in matrix form as
\begin{equation}
\mathbf{F}
=
\left[
\left(-\Omega^2+\omega^2+\frac{i\Omega}{\tau_p}\right)\mathbb{I}_2
- g\bar{S}\Omega\,\sigma_y
\right]\mathbf{Q}.
\end{equation}

We define
\begin{equation}
\alpha \equiv -\Omega^2+\omega^2+\frac{i\Omega}{\tau_p},
\qquad
\beta \equiv -\Omega g \chi_p,
\end{equation}
and use the equilibrium spin polarization
$S_{\mathrm{eq}}=\chi_p \bar{L}_z$.
The force then becomes
\begin{equation}
\mathbf{F}
=
\left[
\alpha\,\mathbb{I}_2
+ \beta \bar{L}_z\,\sigma_y
\right]\mathbf{Q}.
\end{equation}

Introducing the matrix
\begin{equation}
\mathbb{H}
=
\begin{pmatrix}
\alpha & -i\beta \bar{L}_z\\
i\beta \bar{L}_z & \alpha
\end{pmatrix},
\end{equation}
its inverse is
\begin{equation}
\mathbb{H}^{-1}
=
\frac{1}{\alpha^2-(\beta\bar{L}_z)^2}
\left(
\alpha\mathbb{I}_2-\beta\bar{L}_z\sigma_y
\right).
\end{equation}
The phonon amplitude satisfies $\mathbf{Q}=\mathbb{H}^{-1}\mathbf{F}$. We take a circularly polarized drive $\mathbf{F}= F\, (1, -i)\, e^{i \Omega t}$, which yields the scalar relation
\begin{equation}
\label{eq:Qbar_solution}
\bar{Q}
=
\frac{F}{\alpha-\beta \bar{L}_z}.
\end{equation}

\subsubsection{Phonon angular momentum and self-consistent equation}
The phonon angular momentum associated with the circular motion is
\begin{equation}
\bar{L}_z
=
\text{Im}\left(\psi^*\dot{\psi}\right)
=
\Omega|\bar{Q}|^2, \quad \text{where } \psi = |\bar{Q}| e^{i (\Omega t + \theta)},
\end{equation}
which leads to the self-consistent equation
\begin{equation}
\label{eq:self_consistent_Lz}
\bar{L}_z
=
\Omega
\left|
\frac{F}{\alpha-\beta \bar{L}_z}
\right|^2.
\end{equation}
Rearranging, this can be written as
\begin{equation}
\bar{L}_z\left|\alpha-\beta \bar{L}_z\right|^2
-\Omega F^2
=0.
\end{equation}

The steady-state phonon angular momentum is determined by the zeros of the function
\begin{equation}
f(\bar{L}_z)
\equiv
\bar{L}_z\left|\alpha-\beta \bar{L}_z\right|^2
-\Omega F^2.
\end{equation}

For convenience, from now on we define $A \equiv |\bar{Q}|$ (both in the main text and in this Supplemental Material).
\vspace{0.5cm}

\subsection{Linearization around the steady state solution}
\label{subsec:linearization}

To analyze the stability of the steady-state solution, we linearize the dynamics around the periodic orbit $\bm{Q}'(t)$ and the corresponding spin value $S'$. We introduce perturbations
\begin{equation}
\bm{Q}(t)=\bm{Q}'(t)+\bm{\eta}(t),
\qquad
S(t)=S'+\zeta(t),
\end{equation}
and retain only terms linear in $(\bm{\eta},\zeta)$. Defining
\begin{equation}
\bm{\Psi}(t)=\big(\eta_x,\eta_y,\dot{\eta}_x,\dot{\eta}_y,\zeta\big)^{\mathrm T},
\end{equation}
the linearized dynamics can be written as
\begin{align}
\label{eq:stability_ode}
\frac{d}{dt}\bm{\Psi}(t)=\mathbb{J}(t)\,\bm{\Psi}(t),
\end{align}
where $\mathbb{J}(t)$ is the $5\times 5$ non-Hermitian matrix
\begin{equation}
\label{eq:H_matrix_fixedwidth}
\begin{minipage}{15cm} 
\centering
\resizebox{\linewidth}{!}{$
\mathbb{J}(t)=
\begin{pmatrix}
0 & 0 & 1 & 0 & 0 \\
0 & 0 & 0 & 1 & 0 \\
-\omega^2 - \dfrac{g \, P^{\prime}_y(t) \, Q^{\prime}_y(t) \, \chi_p}{2 \, \tau_s} &
\dfrac{g \, P^{\prime}_x(t) \, Q^{\prime}_y(t) \, \chi_p}{2 \, \tau_s} &
-\dfrac{1}{\tau_p} + \dfrac{g \, {Q^{\prime}_y(t)}^2 \, \chi_p}{2 \, \tau_s} &
-\dfrac{g \, Q^{\prime}_x(t) \, Q^{\prime}_y(t) \, \chi_p}{2 \, \tau_s} - g \, S^{\prime} &
-g \, P^{\prime}_y(t) + \dfrac{g \, Q^{\prime}_y(t)}{2 \, \tau_s} \\
\dfrac{g \, P^{\prime}_y(t) \, Q^{\prime}_x(t) \, \chi_p}{2 \, \tau_s} &
-\omega^2 - \dfrac{g \, P^{\prime}_x(t) \, Q^{\prime}_x(t) \, \chi_p}{2 \, \tau_s} &
-\dfrac{g \, Q^{\prime}_x(t) \, Q^{\prime}_y(t) \, \chi_p}{2 \, \tau_s} + g \, S^{\prime} &
-\dfrac{1}{\tau_p} + \dfrac{g \, {Q^{\prime}_x(t)}^2 \, \chi_p}{2 \, \tau_s} &
g \, P^{\prime}_x(t) - \dfrac{g \, Q^{\prime}_x(t)}{2 \, \tau_s} \\
\dfrac{P^{\prime}_y(t) \, \chi_p}{\tau_s} &
-\dfrac{P^{\prime}_x(t) \, \chi_p}{\tau_s} &
-\dfrac{Q^{\prime}_y(t) \, \chi_p}{\tau_s} &
\dfrac{Q^{\prime}_x(t) \, \chi_p}{\tau_s} &
-\dfrac{1}{\tau_s}
\end{pmatrix}
$}
\end{minipage}
\end{equation}
Here $Q'_{x,y}(t)$ and $P'_{x,y}(t)$ denote the steady state solution and its first derivative respectively. The explicit time dependence enters only through these periodic functions. This matrix is periodic, with the same periodicity as the driving field, such that \(\mathbb{J}(t) = \mathbb{J}(t + T_0)\).

\subsection{Floquet criterion for stability}
\label{subsec:floquet}

Eq.~\eqref{eq:stability_ode} is a linear system with periodic coefficients, so its stability is governed by Floquet theory. Let $\mathds{U}(t)$ be the fundamental matrix solution of
\begin{equation}
\label{eq:fundamental_matrix}
\frac{d}{dt}\mathds{U}(t)=\mathbb{J}(t)\,\mathds{U}(t).
\end{equation}
The matrix
\begin{equation}
\label{eq:monodromy}
\mathds{M}\equiv \mathds{U}(T_0)
\end{equation}
propagates perturbations by one drive period:
\begin{equation}
\bm{\Psi}(T_0)=\mathds{M}\,\bm{\Psi}(0).
\end{equation}
Equivalently,
\begin{equation}
\label{eq:monodromy_matrix}
\mathds{M}=\hat{T}\exp\!\left(\int_{0}^{T_0}\!dt\,\mathbb{J}(t)\right),
\end{equation}
where $\hat{T}$ denotes time ordering.

Let $\{\mu_j\}$ be the eigenvalues of $\mathds{M}$. The associated Floquet exponents are
\begin{equation}
\label{eq:floquet_exponents}
\lambda_j=\frac{1}{T_0}\ln \mu_j.
\end{equation}
Stability of the steady state solution requires that all perturbations decay, which is equivalent to either of the following conditions:
\begin{align}
\label{eq:stability_criteria}
|\mu_j|<1 \ \text{for all } j
\qquad \Longleftrightarrow \qquad
\text{Re}(\lambda_j)<0 \ \text{for all } j.
\end{align}

In our problem, crossing this threshold marks the loss of stability of the steady state and the emergence of a new dynamical regime in which additional frequencies appear.

\subsection{Numerical evaluation of the Floquet spectrum}
\label{subsec:numerics_floquet}

We discretize $[0,T_0]$ into $N$ steps of size $\delta t=T_0/N$ with $t_k=k\delta t$. Approximating $\mathbb{J}(t)$ as piecewise constant on each interval gives
\begin{equation}
\label{eq:discrete_product}
\mathds{M}\approx \prod_{k=0}^{N-1}\exp\!\big(\mathbb{J}(t_k)\,\delta t\big),
\end{equation}
where the product is ordered in increasing time. Diagonalizing the resulting $\mathds{M}$ yields the Floquet multipliers.

\begin{figure}[t] 
\centering \includegraphics[width=6cm]{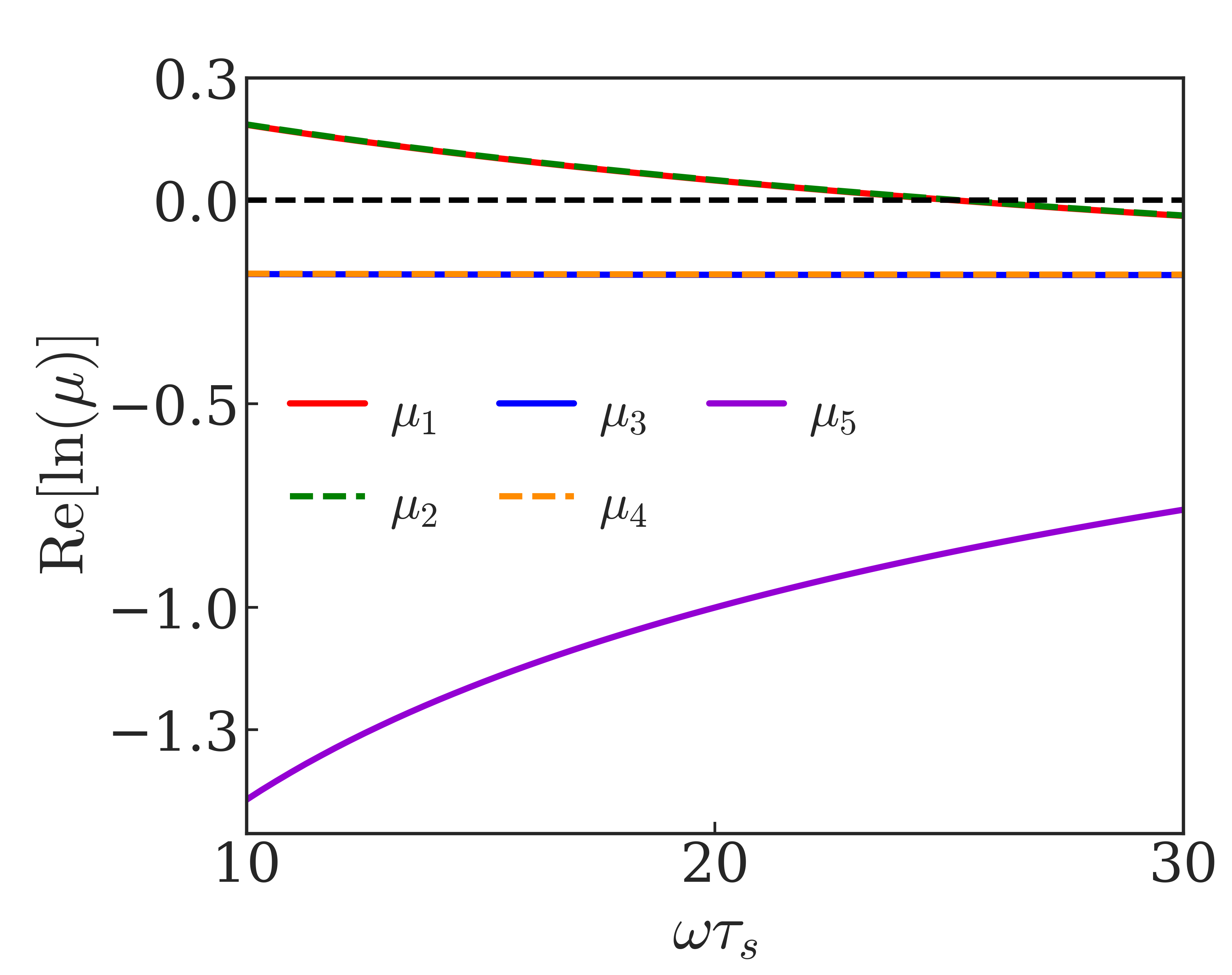} \caption{Stability analysis: real part of the Floquet exponents, $\text{Re}[\ln \mu]$, as a function of $\tau_s$, indicating stability when all values are negative. The parameters used are: $\chi_p = 4.0/\hbar$, $g=\chi_p(10/1.16^2)\hbar\omega$, $\Omega = 1.16\omega$, $F = 0.08\omega\sqrt{\hbar \omega}$ and $\omega\tau_p = 10$.} 
\label{fig:floquet} 
\end{figure}

We numerically calculate \(\mu\). The real part of $\ln{\mu}$ is plotted in Fig.~\ref{fig:floquet} as a function of the spin relaxation time $\tau_s$. At large $\tau_s$, the real parts of $\ln{\mu}$ are all negative, indicating that the amplitude of the fluctuations $\Psi$ decays exponentially. The limit-cycle state is thus stable. As the $\tau_s$ reduces, two of $\ln\mu$ show a positive real part, indicating that the fluctuations increase exponentially as time evolves. Thus, the limit-cycle state becomes unstable, and a dynamical transition to a different steady state happens.

\section{Dissipation-induced instability and emergence of temporal order}

In this section, we derive an explicit retarded phonon equation by eliminating the
spin variable and show how spin relaxation produces a phase-lagged feedback that
can act as an effective anti-damping, destabilizing the time-locked state and
enabling the breaking of discrete time-translation symmetry.

\subsection{Equations of motion}

\subsubsection{Phonon sector}
We start from the phonon Lagrangian:
\begin{equation}
L_{\rm ph}
=
\frac{1}{2}\dot{\Q}^{\,2}
-\frac{\omega^2}{2}\Q^2
-\frac{g}{2}\,S\,\z\cdot(\Q\times\dot{\Q})
+\bm{F}(t)\cdot\Q,
\label{eq:Lph}
\end{equation}
which yields the phonon equation of motion:
\begin{equation}
\ddot{\Q}
=
-\omega^2\Q+\bm{F}(t)
-g\,\dot{\Q}\times(S\,\z)
-\frac{g}{2}\,\Q\times(\dot{S}\,\z)
-\frac{1}{\tau_p}\dot{\Q}.
\label{eq:Qeom}
\end{equation}
The phonon angular momentum is:
\begin{equation}
L_z(t)=\z\cdot(\Q\times\dot{\Q}).
\label{eq:lz_def}
\end{equation}

\subsubsection{Spin sector (relaxational Bloch equation)}
In the paramagnetic regime, spin obeys the Bloch equation as shown in Eq.\eqref{eq:SM_bloch_from_lindblad} and Eq.\eqref{eq:SM_seq_lz}:
\begin{equation}
\dot{S}(t)
=
-\frac{1}{\tau_s}\Big(S(t)-S_{\rm eq}(L_z(t))\Big),
\qquad
S_{\rm eq}(L_z)=\tanh(\chi_p L_z).
\label{eq:Bloch}
\end{equation}

\subsection{Exact elimination of \(S(t)\) and consistent treatment of \(\dot S(t)\)}

Equation \eqref{eq:Bloch} is a linear ODE in \(S(t)\) with a time-dependent source \(S_{\rm eq}(L_z(t))\). Its steady causal solution is
\begin{equation}
S(t)=\int_{-\infty}^{t}\frac{d t'}{\tau_s}\,e^{-(t-t')/\tau_s}\,S_{\rm eq}(L_z(t')).
\label{eq:Sconv_exact}
\end{equation}
Importantly, \(\dot S(t)\) must be treated consistently. It follows directly from \eqref{eq:Bloch}:
\begin{equation}
\dot{S}(t)
=
-\frac{1}{\tau_s}S(t)+\frac{1}{\tau_s}S_{\rm eq}(L_z(t)).
\label{eq:Sdot_exact}
\end{equation}
Substituting \eqref{eq:Sconv_exact} into \eqref{eq:Sdot_exact} gives an explicit decomposition into an instantaneous
term and a retarded term:
\begin{equation}
\dot{S}(t)
=
\frac{1}{\tau_s}S_{\rm eq}(L_z(t))
-\int_{-\infty}^{t}\frac{d t'}{\tau_s^2}\,e^{-(t-t')/\tau_s}\,S_{\rm eq}(L_z(t')).
\label{eq:Sdot_inst_ret}
\end{equation}

\subsection{Linearization near the time-locked state}

Near the onset of instability, we expand
\begin{equation}
S_{\rm eq}(L_z)=\chi_p L_z-\frac{\chi_p^3}{3}L_z^3+\cdots,
\label{eq:Seq_expand}
\end{equation}
and keep only the linear term to obtain the leading-order feedback responsible for the threshold:
\begin{align}
S(t) &=
\chi_p\int_{-\infty}^{t}\frac{d t'}{\tau_s}\,e^{-(t-t')/\tau_s}\,L_z(t') +\mathcal{O}\!\left(\chi_p^3\,L_z^3\right)\label{eq:S_lin}
\\
\dot{S}(t) & =
\frac{\chi_p}{\tau_s}L_z(t)
-\chi_p\int_{-\infty}^{t}\frac{d t'}{\tau_s^2}\,e^{-(t-t')/\tau_s}\,L_z(t') +\mathcal{O}\!\left(\chi_p^3 \,L_z^3 \right).
\label{eq:Sdot_lin}
\end{align}
Cubic and higher terms in \eqref{eq:Seq_expand} provide nonlinear saturation beyond threshold and can be reinstated
systematically.

\subsection{Retarded phonon equation (including the \(\dot S\) term)}

Substituting \eqref{eq:S_lin} and \eqref{eq:Sdot_lin} into \eqref{eq:Qeom} yields the properly reduced retarded phonon dynamics:
\begin{align}
\ddot{\Q}(t)
&=
-\omega^2\Q(t)+\bm{F}(t)-\frac{1}{\tau_p}\dot{\Q}(t)
\nonumber\\
&\quad
-g\chi_p\,\dot{\Q}(t)\times\z
\int_{-\infty}^{t}\frac{d t'}{\tau_s}e^{-(t-t')/\tau_s}\,L_z(t')
-\frac{g\chi_p}{2\tau_s}\,\Q(t)\times\z\,L_z(t)
+\frac{g\chi_p}{2}\,\Q(t)\times\z
\int_{-\infty}^{t}\frac{d t'}{\tau_s^2}e^{-(t-t')/\tau_s}\,L_z(t')
\nonumber\\
&\quad
+\mathcal{O}\!\left(\chi_p^3\, L_z^3 \right).
\label{eq:Q_retarded}
\end{align}

The \(S\)-term produces a retarded Lorentz-like feedback with kernel \(K_1(t)=\tau_s^{-1}e^{-t/\tau_s}\).
The \(\dot S\)-term produces:\\
(i) An instantaneous contribution \(-\frac{g\chi_p}{2\tau_s}\Q\times\z\,L_z(t)\).\\
(ii) An additional retarded feedback with kernel \(K_2(t)=\tau_s^{-2}e^{-t/\tau_s}\).

\subsection{Frequency-domain interpretation (phase lag and anti-damping)}

For a Fourier component $L_z(t)\sim e^{i\omega t}$, the linear response implied by \eqref{eq:S_lin} gives
\begin{equation}
S(\omega)=\chi_p\frac{1}{1+i\omega\tau_s}\,L_z(\omega)
+\mathcal{O}\!\left(\chi_p^3\,[L_z^3](\omega)\right),
\label{eq:S_omega_O}
\end{equation}
and therefore
\begin{equation}
\dot S(\omega)= i\omega S(\omega)
=\chi_p\frac{i\omega}{1+i\omega\tau_s}\,L_z(\omega)
+\mathcal{O}\!\left(\chi_p^3\,[L_z^3](\omega)\right).
\label{eq:Sdot_omega_O}
\end{equation}
Here $[L_z^3](\omega)$ denotes the Fourier transform of $L_z(t)^3$. Using this, the spin response is characterized by a complex susceptibility, whose
imaginary part introduces a phase lag in the oscillation of the angular momentum.

As can be seen from Eq.~\eqref{eq:Qeom}, the spin--phonon coupling generates two
contributions to the force acting on the phonon coordinates:
\begin{equation}
\bm F_S = - g\,\dot{\bm Q}\times (S\hat z)
-\frac{g}{2}\,\bm Q\times (\dot S\,\hat z).
\end{equation}

Since the instantaneous power is defined as $P(t)=\bm F_S\cdot\dot{\bm Q}$, only
the second term contributes to the power, as the first one is always perpendicular
to $\dot{\bm Q}$. Therefore,
\begin{equation}
P(t)=\bm F_S\cdot\dot{\bm Q}
= -\frac{g}{2}\,(\bm Q\times (\dot S\,\hat z))\cdot\dot{\bm Q}
=\frac{g}{2}\,\dot S(t)\,L_z(t),
\end{equation}
where $L_z=\hat z\cdot(\bm Q\times\dot{\bm Q})$.
The work performed per drive period is then defined as
\begin{equation}
W=\int_{t}^{t+T_0} \, P(t) \,dt.
\end{equation}

The time derivative in $\dot S$ introduces a phase shift of $\pi/2$ relative to $S$. Thus, if $S$ and $L_z$ oscillate in phase, then $\dot S$ is shifted by $\pi/2$ relative to $L_z$, and the net work performed over one period vanishes. However, the imaginary part of the susceptibility induces an additional phase lag between $S$ and $L_z$, which in turn yields a finite phase correlation between $\dot S$ and $L_z$. Consequently, the cycle-averaged work becomes nonzero, providing an effective anti-damping mechanism for the phonon orbit.

\subsection{Exact ODE embedding (extended Markovian system)}

Equation \eqref{eq:Q_retarded} contains exponential memory kernels. Therefore, the \emph{linear} retarded terms
can be embedded exactly into a finite-dimensional ODE system by introducing the memory variables
\begin{align}
m_1(t) &\equiv \int_{-\infty}^{t}\frac{d t'}{\tau_s}e^{-(t-t')/\tau_s}\,L_z(t'),
\label{eq:m1_def}
\\
m_2(t) &\equiv \int_{-\infty}^{t}\frac{d t'}{\tau_s^2}e^{-(t-t')/\tau_s}\,L_z(t')=\frac{m_1(t)}{\tau_s}.
\label{eq:m2_def}
\end{align}
Differentiation under the integral sign gives the exact ODEs
\begin{equation}
\dot m_1(t)=-\frac{1}{\tau_s}m_1(t)+\frac{1}{\tau_s}L_z(t),
\qquad
\dot m_2(t)=-\frac{1}{\tau_s}m_2(t)+\frac{1}{\tau_s^2}L_z(t)=\frac{\dot m_1(t)}{\tau_s}.
\label{eq:m_odes}
\end{equation}
Using \eqref{eq:m1_def}--\eqref{eq:m_odes}, the equation of motion \eqref{eq:Q_retarded} can be written as
\begin{align}
\ddot{\Q}(t)&=
-\omega^2\Q(t)+\bm{F}(t)-\frac{1}{\tau_p}\dot{\Q}(t)
-g\chi_p\,\dot{\Q}(t)\times\z\,m_1(t)
-\frac{g\chi_p}{2\tau_s}\Q(t)\times\z\,(L_z(t)-m_1(t)) +\mathcal{O}\!\left(\chi_p^3\,L_z^3\right),
\label{eq:Q_extended}
\end{align}
with \(L_z=\z\cdot(\Q\times\dot{\Q})\). Equations \eqref{eq:Q_extended} and \eqref{eq:m_odes} form a closed finite-dimensional (periodically forced) ODE system for the linear retarded feedback.

\subsection{Rigorous linear stability framework (Floquet analysis)}
We analyze the stability of the reduced extended system using the same Floquet framework introduced in \ref{subsec:floquet} and \ref{subsec:numerics_floquet}, now applied to the memory-embedded formulation.
Let \(\Q_0(t)\) be a time-locked steady solution with the drive period \(T_0=2\pi/\Omega\):
\(\Q_0(t+T_0)=\Q_0(t)\).
Define the state vector
\begin{equation}
\bm{x}(t)=
\begin{pmatrix}
\Q(t)\\ \dot{\Q}(t)\\ m_1(t)
\end{pmatrix},
\qquad
\bm{x}_0(t)=
\begin{pmatrix}
\Q_0(t)\\ \dot{\Q}_0(t)\\ m_{1,0}(t)
\end{pmatrix}.
\label{eq:xvec}
\end{equation}
A perturbation \(\Omega_s\bm{x}\) obeys
\begin{equation}
\Omega_s\dot{\bm{x}}(t)=\mathbb{J}(t)\,\Omega_s\bm{x}(t),
\qquad
\mathbb{J}(t+T_0)=\mathbb{J}(t),
\label{eq:linFloquet}
\end{equation}
where \(\mathbb{J}(t)\) is the Jacobian evaluated on \(\bm{x}_0(t)\). Stability is controlled by the monodromy matrix defined in Eq.\eqref{eq:monodromy_matrix}, whose eigenvalues \(\{\mu_\alpha\}\) are the Floquet multipliers. The orbit is stable if $\mu_\alpha$ satisfy the criteria established in Eq.\eqref{eq:stability_criteria}, i.e. \(|\mu_\alpha|<1\) for all non-neutral multipliers.

A dissipative transition to a non-\(T_0\)-periodic attractor occurs when a complex pair crosses the unit circle:
\begin{equation}
\mu_\pm=e^{\pm i\Omega_s T_0},
\qquad
|\mu_\pm|=1 \text{ at threshold}.
\label{eq:mu_pair}
\end{equation}
This produces a secondary frequency \(\Omega_s\) not fixed by the drive. If \(\Omega_s/\Omega\) is irrational the state is quasiperiodic; if rational it yields a higher-period orbit.

\subsection{Controlled analytical reduction near threshold: sideband amplitude equation for \(B(t)\)}
\label{subsection_H}

\subsubsection*{Complex coordinate and scalar form of the spin feedback}
Introduce the circular complex coordinate
\begin{equation}
\psi(t)\equiv Q_x(t)+iQ_y(t),
\qquad
\dot\psi(t)=\dot Q_x(t)+i\dot Q_y(t).
\label{eq:psi_def}
\end{equation}
The angular momentum is
\begin{equation}
L_z(t)=\im\!\big(\psi^*(t)\dot\psi(t)\big)
=\frac{1}{2i}\Big(\psi^*\dot\psi-\psi\,\dot\psi^{\,*}\Big).
\label{eq:lz_psi}
\end{equation}
In the complex representation, the spin terms in \eqref{eq:Q_extended} become a scalar feedback of the form
\begin{align}
\ddot\psi+\frac{1}{\tau_p}\dot\psi+\omega^2\psi
&=
F(t)
+i g\chi_p\,\dot\psi\,m_1
+i\frac{g\chi_p}{2\tau_s}\psi\,(L_z-m_1) +\mathcal{O}\!\left(
\chi_p^3\,L_z^3
\right).
\label{eq:main_eom}
\end{align}

where $F(t) = F_0 e^{i\Omega t}$. We assume a known steady-state solution $\psi_0 = A e^{i\Omega t}$ where $A$, $L_z$, and $m_1$ are constants. Let the constant steady-state values for angular momentum and the memory term be $L_0$. From the definition of $L_z$:
\begin{equation}
    L_0 = \text{Im}(A^* e^{-i\Omega t} \cdot (i\Omega A) e^{i\Omega t}) = \text{Im}(i\Omega |A|^2) = \Omega |A|^2
\end{equation}
Since $\dot{m}_1=0$ in the steady state, $m_1 = L_z = L_0$.

\subsubsection*{Linearization}

We introduce a trial solution $\psi(t) = (A + B(t)) e^{i\Omega t}$, where $|B| \ll A$.
The derivatives are:
\begin{align}
    \dot{\psi} &= (\dot{B} + i\Omega(A+B))e^{i\Omega t} \\
    \ddot{\psi} &= (\ddot{B} + 2i\Omega\dot{B} - \Omega^2(A+B))e^{i\Omega t}
\end{align}

\subsubsection*{Perturbation of \texorpdfstring{$L_z$ and $m_1$}{lz and m1}}
Substituting the ansatz into $L_z$ and retaining only terms linear in $B$:
\begin{align}
    L_z &= \text{Im}\left[ (A^*+B^*)e^{-i\Omega t} \cdot (\dot{B} + i\Omega(A+B))e^{i\Omega t} \right] \nonumber \\
    &= \text{Im}\left[ A^*\dot{B} + i\Omega|A|^2 + i\Omega(A^*B + AB^*) \right] +  \text{Im}\left[B^* \dot{B} +i \Omega |B|^2 \right] \nonumber \\
    &=\underbrace{L_0}_{ \Omega |A|^2} + \underbrace{\text{Im}(A^*\dot{B}) + \Omega(A^*B + AB^*)}_{\delta L_z(t)} +\mathcal{O}\!\left(\Omega|B|^2\right).
\end{align}
The perturbation $\delta m_1$ is related to $\delta L_z$ by the integral solution:
\begin{equation}
    \delta m_1(t) = \int_{-\infty}^{t} \frac{dt'}{\tau_s} e^{-(t-t')/\tau_s} \delta L_z(t')
\end{equation}

\subsubsection*{Perturbation of the Equation of Motion}
Substituting $\psi$, $\dot{\psi}$, $\ddot{\psi}$, $L_z = L_0 + \delta L_z$, and $m_1 = L_0 + \delta m_1$ into Eq.~\eqref{eq:main_eom}, and subtracting the zeroth-order equation satisfied by $A$, we obtain the equation for $B$.

The Left Hand Side (LHS) becomes:
\begin{equation}
    \text{LHS} = \left[ \ddot{B} + \left(\frac{1}{\tau_p} + 2i\Omega\right)\dot{B} + \left(\omega^2 - \Omega^2 + \frac{i\Omega}{\tau_p}\right)B \right] e^{i\Omega t}
\end{equation}

The Right Hand Side (RHS) terms are linearized as follows:
\begin{enumerate}
    \item 
    \begin{align*}
        i g \chi_p \dot{\psi} m_1 &= i g \chi_p \left[ (\dot{B} + i\Omega B)L_0 + i\Omega A\delta m_1 \right] e^{i\Omega t} + i g \chi_p \left[ (\dot{B} + i\Omega B)\delta m_1 \right] e^{i\Omega t}\\
        &=i g \chi_p \left[ (\dot{B} + i\Omega B)L_0 + i\Omega A\delta m_1 \right] e^{i\Omega t} +\mathcal{O}\!\left(e^{i\Omega t}\chi_p|B|^2\right)
    \end{align*}
   
    \item 
    \begin{align*}
        i \frac{g\chi_p}{2\tau_s} \psi (L_z - m_1) &= i \frac{g\chi_p}{2\tau_s} A (\delta L_z - \delta m_1) e^{i\Omega t} + i \frac{g\chi_p}{2\tau_s} B (\delta L_z - \delta m_1) e^{i\Omega t}\\ 
        &= i \frac{g\chi_p}{2\tau_s} A (\delta L_z - \delta m_1) e^{i\Omega t} +\mathcal{O}\!\left(e^{i\Omega t}\chi_p|B|^2\right)
    \end{align*}

    \item 
   Here we must also take into account how the correction change and subtract the zero-order terms from them.

   \begin{align}
       \mathcal{O}\!\left(\chi_p^3\,L_z^3
\right) = \mathcal{O}(e^{i\Omega t}i g \chi_p^3|B|) +\mathcal{O}(e^{i\Omega t}\chi_p^3|B|^2)
\label{eq_correction_linear}
   \end{align}

    \end{enumerate}

\subsection{Final Equation for $B(t)$}

Equating LHS and RHS, canceling $e^{i\Omega t}$, and grouping terms involving $B$ on the left:

\begin{equation}
    \ddot{B} + \Gamma_{\text{eff}} \dot{B} + \Omega_{\text{eff}}^2 B = \mathcal{S}(B, B^*)+ \mathcal{O}( g \chi_p^3|B|) +\mathcal{O}(\chi_p^3|B|^2) + \mathcal{O}\!\left(\chi_p|B|^2\right)
    \label{eq:B_eom}
\end{equation}

The term $\mathcal{O}( g \chi_p^3|B|)$ comes from the $\mathcal{O}(\chi_p^3 L_z^3)$ correction (Eq. \eqref{eq_correction_linear}) and it is therefore correctly neglected in our truncation, even if it is linear in $|B|$. Here, the effective damping and frequency coefficients are:
\begin{align}
\label{eq:effective_parameters}
    \Gamma_{\text{eff}} &= \frac{1}{\tau_p} + i(2\Omega - g\chi_p L_0) \quad \text{and} \notag\\
    \Omega_{\text{eff}}^2 &= \omega^2 - \Omega^2 + i\Omega \left( \frac{1}{\tau_p} -ig\chi_p L_0 \right),
\end{align}
and the spin-feedback source term $\mathcal{S}$ is:
\begin{equation}
\label{eq:S}
    \mathcal{S}(B, B^*) = -\left( g\chi_p \Omega A + \frac{i g\chi_p A}{2\tau_s} \right) \delta m_1(t) + \frac{i g\chi_p A}{2\tau_s} \delta L_z(t)
\end{equation}
with $\delta L_z$ and $\delta m_1$ defined as functions of $B$ above.

\subsection{Derivation of the Characteristic Equation for $\Omega_s$}

We start with the linearized equation of motion for the perturbation $B(t)$ derived previously:
\begin{equation}
    \ddot{B} + \Gamma_{\text{eff}} \dot{B} + \Omega_{\text{eff}}^2 B = \left( \lambda_1 \delta m_1(t) + \lambda_2 \delta L_z(t) \right)
    \label{eq:B_linear}
\end{equation}
where we have defined the source coefficients for compactness:
\begin{align}
    \lambda_1 &= -g\chi_p A \left( \Omega + \frac{i}{2\tau_s} \right) \\
    \lambda_2 &= i \frac{g\chi_p A}{2\tau_s}
\end{align}
and the effective parameters are given by \ref{eq:effective_parameters}.

\subsubsection*{1. Ansatz and Spectral Decomposition}
We assume the trial solution:
\begin{equation}
\label{eq:B}
    B(t) = B_0 e^{i\Omega_s t}
\end{equation}
where $B_0$ is a complex constant and $\Omega_s$ is the complex mode frequency. Consequently, the complex conjugate is:
\begin{equation}
    B^*(t) = B_0^* e^{-i\Omega_s^* t}
\end{equation}
Substituting the time derivative $\dot{B} = i\Omega_s B$, the Left Hand Side (LHS) of Eq.~\eqref{eq:B_linear} becomes:
\begin{equation}
    \text{LHS} = \left[ -\Omega_s^2 + i\Omega_s \Gamma_{\text{eff}} + \Omega_{\text{eff}}^2 \right] B_0 e^{i\Omega_s t}
    \label{eq:LHS_ansatz}
\end{equation}

\subsubsection*{2. Spectral Form of $\delta L_z$ and $\delta m_1$}
Substituting the ansatz into the linearized angular momentum expression:
\begin{align}
    \delta L_z(t) &= \text{Im}(A^*\dot{B}) + \Omega(A^*B + AB^*) \nonumber \\
    &= \text{Im}(i\Omega_s A^* B_0 e^{i\Omega_s t}) + \Omega(A^* B_0 e^{i\Omega_s t} + A B_0^* e^{-i\Omega_s^* t}) \nonumber \\
    &= \left( \frac{\Omega_s}{2} + \Omega \right) A^* B_0 e^{i\Omega_s t} + \left( \frac{\Omega_s^*}{2} + \Omega \right) A B_0^* e^{-i\Omega_s^* t}
\end{align}
Let $\kappa(\Omega_s) = \left( \Omega + \frac{\Omega_s}{2} \right)$. We can write:
\begin{equation}
    \delta L_z(t) = \kappa(\Omega_s) A^* B_0 e^{i\Omega_s t} + \kappa(\Omega_s^*) A B_0^* e^{-i\Omega_s^* t}
\end{equation}

The memory term $\delta m_1$ is the solution to $\dot{\delta m_1} + \frac{1}{\tau_s}\delta m_1 = \frac{1}{\tau_s}\delta L_z$. This is a linear filter operation. Defining the susceptibility $\chi_s(\omega) = (1 + i\omega\tau_s)^{-1}$, we integrate each frequency component:
\begin{equation}
\label{eq:delta_m1_w}
    \delta m_1(t) = \frac{\kappa(\Omega_s) A^* B_0}{1 + i\Omega_s \tau_s} e^{i\Omega_s t} + \frac{\kappa(\Omega_s^*) A B_0^*}{1 - i\Omega_s^* \tau_s} e^{-i\Omega_s^* t}
\end{equation}

\subsubsection*{3. The Characteristic Equation for $\Omega_s$}
We substitute $\delta L_z$ and $\delta m_1$ into the RHS of Eq.~\eqref{eq:B_linear}. The equation must hold for the coefficients of the basis function $e^{i\Omega_s t}$.

\textbf{Right Hand Side}($e^{i\Omega_s t}$ component):
The terms proportional to $e^{i\Omega_s t}$ come from the first terms of $\delta L_z$ and $\delta m_1$:
\begin{align}
    \text{RHS}|_{e^{i\Omega_s t}} &= \left[ \lambda_1 \frac{\kappa(\Omega_s) A^*}{1 + i\Omega_s \tau_s} + \lambda_2 \kappa(\Omega_s) A^* \right] B_0 \nonumber \\
    &= \kappa(\Omega_s) A^* \left[ \frac{\lambda_1}{1 + i\Omega_s \tau_s} + \lambda_2 \right] B_0
\end{align}

\begin{figure}[h!] 
\centering \includegraphics[width=8cm]{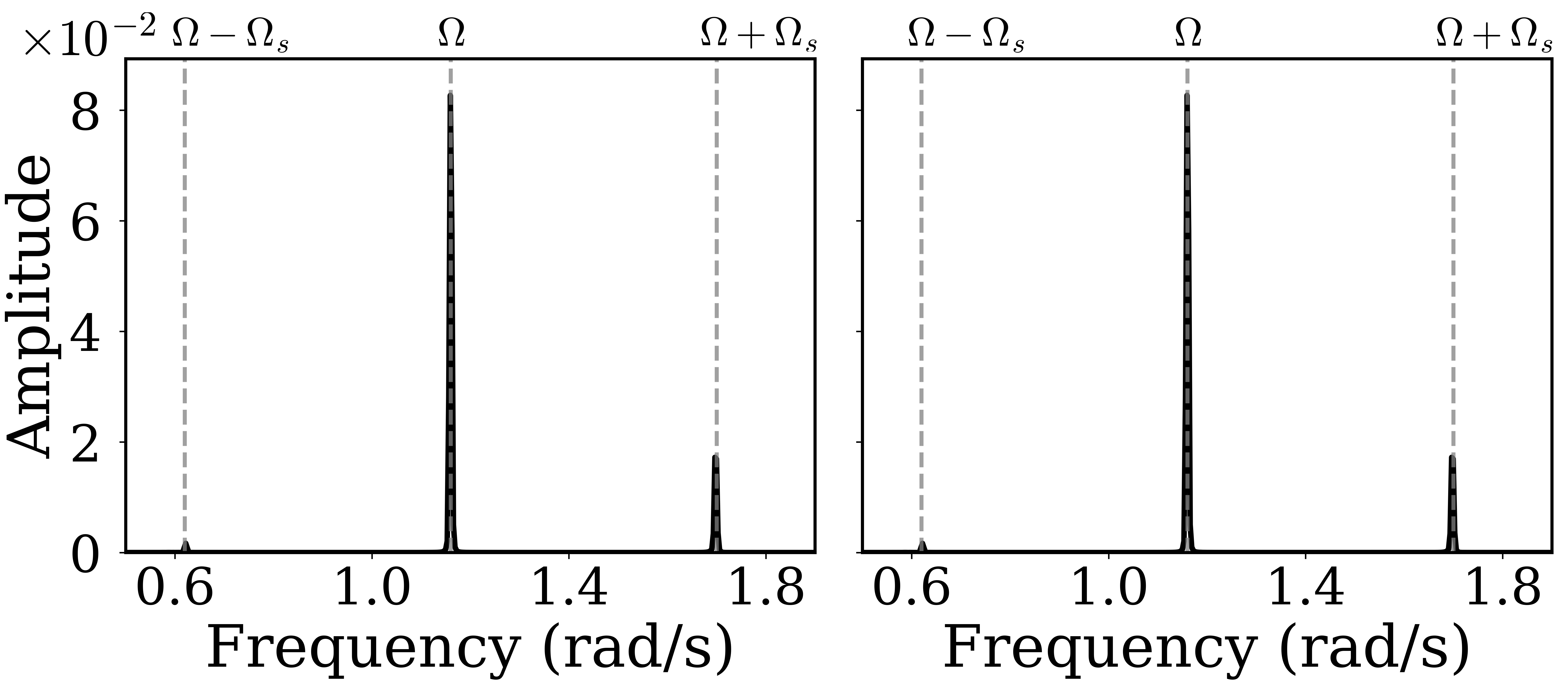}  \caption{Fourier spectra of the phonon coordinates. Left panel, $Q_x(t)$, and right panel, $Q_y(t)$. The parameters used are $\Omega = 1.16\omega$, $F = 0.08\omega\sqrt{\hbar \omega}$, $\chi_p = 4.0/\hbar$, $g = \chi_p(10/1.16^2)\hbar\omega$, $\omega\tau_p = 10$ and $\omega\tau_s=19$.
}
\label{fig:fft_Q} 
\end{figure}

\textbf{Right Hand Side} ($e^{-i\Omega_s^* t}$ component coupling): 
Note that the equation also generates terms oscillating as $e^{-i\Omega_s^* t}$ proportional to $B_0^*$. In general, obtaining a closed scalar equation requires either solving the coupled system for $(B_0,B_0^*)$ or adopting the rotating-wave approximation (discarding counter-rotating contributions). 
Here, the rotating-wave approximation is supported by the numerical spectra of the phonon coordinates: the Fourier transforms of $Q_x(t)$ and $Q_y(t)$ show a pronounced asymmetry between the sidebands around the drive, with the co-rotating sideband at $\Omega+\Omega_s$ carrying substantially more spectral weight than the counter-rotating sideband at $\Omega-\Omega_s$ (Fig.~\ref{fig:fft_Q}). 
This indicates that the $B_0^*$-driven contribution is subdominant in the regime of interest, and we therefore retain only the resonant terms matching $e^{i\Omega_s t}$, which provide the primary equation for the amplitude $B_0$.

Equating LHS and the resonant part of RHS:
\begin{equation}
\label{eq:Cancel_B0}
    \left( -\Omega_s^2 + i\Omega_s \Gamma_{\text{eff}} + \Omega_{\text{eff}}^2 \right) B_0 = \kappa(\Omega_s) A^* \left( \frac{\lambda_1}{1 + i\Omega_s \tau_s} + \lambda_2 \right) B_0
\end{equation}

\subsubsection*{4. Final Dispersion Relation}
Assuming $B_0 \neq 0$, we divide by $B_0$ to find the equation determining $\Omega_s$:

\begin{equation}
\label{eq:just_omega_s}
    \mathcal{D}(\Omega_s) = -\Omega_s^2 + i\Omega_s \Gamma_{\text{eff}} + \Omega_{\text{eff}}^2 - \mathcal{F}(\Omega_s) = 0
\end{equation}

where the feedback function $\mathcal{F}(\Omega_s)$ is:
\begin{equation}
    \mathcal{F}(\Omega_s) = A^* \left(\Omega + \frac{\Omega_s}{2}\right) \left[ \frac{-g\chi_p A (\Omega + \frac{i}{2\tau_s})}{1 + i\Omega_s \tau_s} + \frac{i g\chi_p A}{2\tau_s} \right]
\end{equation}
This algebraic equation can be solved for the eigenfrequencies $\Omega_s$.

\subsection{Derivation of the Characteristic Equation for $B_0$}

In the previous section, we derived an expression for $\Omega_s$, which enabled us to identify the transitions between different dynamical regimes by analyzing the real and imaginary parts of $\Omega_s$. However, in Eq.~\eqref{eq:Cancel_B0}, all information regarding the magnitude of $B$ is lost.

We therefore now assume that $B_0$ is not sufficiently small to justify neglecting the term $|B_0|^2$. Under this assumption and assuming that $\Omega_s$ is real, the angular momentum is defined as:
$L_z = \text{Im}(\psi^* \dot{\psi}),$ where
\begin{align}
\psi^* \dot{\psi}
&= i \Omega |A|^2 
   + i(\Omega+\Omega_s) |B_0|^2
   + i(\Omega+\Omega_s) A^* B_0 e^{i\Omega_s t}
   + i \Omega A B_0^* e^{-i\Omega_s t}
\end{align}
Then, using the definition of the angular momentum
\begin{align}
\label{eq:angular_mom}
L_z &= \Omega |A|^2 
   + (\Omega+\Omega_s) |B_0|^2 +\text{Im}\left[
   i(\Omega+\Omega_s) A^* B_0 e^{i\Omega_s t}
   + i \Omega A B_0^* e^{-i\Omega_s t}
\right] \nonumber \\
&= \underbrace{\Omega |A|^2 
   + (\Omega+\Omega_s) |B_0|^2}_{L_0} \underbrace{+ A^* B_0 \left(\frac{2\Omega+\Omega_s}{2}\right) e^{i\Omega_s t}
 + A B_0^* \left(\frac{2\Omega+\Omega_s}{2}\right) e^{-i\Omega_s t}}_{\delta L_z}\nonumber\\
 &= L_0 + \delta L_z.
\end{align}

In this context, $\delta L_z$ should not be interpreted as a small quantity; we retain this notation solely for consistency with the convention adopted in the preceding section.

We can therefore express the angular momentum in the following form:
\begin{align}
L_z
&= L_0 + \ell\, e^{i\Omega_s t} + \ell^* e^{-i\Omega_s t},
\end{align}
where 
\begin{align}
 \ell =A^* B_0 \left(\frac{2\Omega+\Omega_s}{2}\right).
\end{align}

Substituting the definition in Eq.\eqref{eq:m_odes} and explicitly writing $m_1$ in the form:
\begin{equation}
m_1 = M_0 + \delta m_1,
\end{equation}
where \(M_0\) is the constant part and \(\delta m_1\) represents the oscillatory contribution, we obtain:
\begin{align}
\delta \dot{m}_1
&= \frac{ \delta L_z -\delta m_1 + (L_0 - M_0)}{\tau_s}.
\end{align}

Since oscillating terms must vanish independently, we require $L_0 = M_0$. Subsequently, we can proceed analogously to the derivation in Eq.~\eqref{eq:delta_m1_w}, defining:
\begin{align}
    \delta m_1 = \alpha e^{i\Omega_s t} + \beta e^{-i\Omega_s t},
\end{align}
and then, obtaining:
\begin{equation}
\delta m_1
= \frac{\ell}{1 + i\Omega_s \tau_s} e^{i\Omega_s t}
+ \frac{\ell^*}{1 - i\Omega_s \tau_s} e^{-i\Omega_s t}.
\end{equation}

Now, we can define $\lambda$ and re-express $\delta L_z$ and $\delta m_1$ as follows: 
\begin{align}
\lambda \;\equiv\; A^{*}\left(\frac{2\Omega+\Omega_s}{2}\right)
\quad\Rightarrow\quad
\delta L_z = \lambda\,B + \lambda^{*}B^{*}.
\end{align}
\begin{align}
\Rightarrow \delta m_1
&= \frac{\lambda}{1+i\Omega_s\tau_s}\,B
+\frac{\lambda^{*}}{1-i\Omega_s\tau_s}\,B^{*}.
\end{align}

By utilizing Eq.\eqref{eq:B_eom} and Eq.\eqref{eq:S}, along with the previously established results, we get:
\begin{align}
\ddot B+\Gamma_{\mathrm{eff}}\dot B+\Omega_{\mathrm{eff}}^{2}B
&=
-\left(g\chi_p\Omega A+\frac{i g\chi_p A}{2\tau_s}\right)
\left(
\frac{\lambda}{1+i\Omega_s\tau_s}\,B
+\frac{\lambda^{*}}{1-i\Omega_s\tau_s}\,B^{*}
\right)
+\frac{i g\chi_p A}{2\tau_s}\left(\lambda B+\lambda^{*}B^{*}\right).
\end{align}

Keeping the contribution proportional to $B$:
\begin{align}
\label{eq:dif_eq_for_B}
\ddot B+\Gamma_{\mathrm{eff}}\dot B+\Omega_{\mathrm{eff}}^{2}B
&=
-\left(g\chi_p\Omega A+\frac{i g\chi_p A}{2\tau_s}\right)
\frac{\lambda}{1+i\Omega_s\tau_s}B
+\frac{i g\chi_p A}{2\tau_s}\lambda B \nonumber
\\
&= \frac{-g\chi_p A\lambda}{1+i\Omega_s\tau_s}\left(\Omega+\frac{\Omega_s}{2}\right)B \nonumber
\\
&= \frac{-g\chi_p}{1+i\Omega_s\tau_s}
\left(\frac{2\Omega+\Omega_s}{2}\right)^2 |A|^2\,B\nonumber\\
&= \mathcal{S}_0 \,B \quad \text{where} \quad \mathcal{S}_0=\frac{-g\chi_p}{1+i\Omega_s\tau_s}
\left(\frac{2\Omega+\Omega_s}{2}\right)^2 |A|^2.
\end{align}

Using the definition in Eq.\eqref{eq:B}:
\begin{align}
B(t) &= B_0 e^{i\Omega_s t}, \quad
\dot B = i\Omega_s B \quad \text{and} \quad
\ddot B = -\Omega_s^2 B.
\end{align}

Therefore, 
\begin{align}
-\Omega_s^2 + i\Omega_s\Gamma_{\mathrm{eff}}+\Omega_{\mathrm{eff}}^{2}
=
\mathcal{S}_0.
\label{eq:master}
\end{align}

As shown in Eq.\eqref{eq:effective_parameters} and for Eq.\eqref{eq:angular_mom}:
\begin{align}
\label{eq_Gamma_full}
\Gamma_{\mathrm{eff}}
&= \frac{1}{\tau_p}+i2\Omega
-i g\chi_p L_0 \nonumber
\\
&= \frac{1}{\tau_p}+i2\Omega
-i g\chi_p\Omega|A|^2
-i g\chi_p(\Omega+\Omega_s)|B|^2\nonumber\\
&= \Gamma_0 + \Gamma^{\prime}|B|^2, \quad \text{where} \quad \Gamma_0= \frac{1}{\tau_p}+i2\Omega
-i g\chi_p\Omega|A|^2 \quad \text{and} \quad \Gamma^{\prime}=-i g\chi_p(\Omega+\Omega_s).
\end{align}
and
\begin{align}
\label{eq_oMEGa_full}
\Omega_{\mathrm{eff}}^{2}
&= \omega^2-\Omega^2+\frac{i\Omega}{\tau_p}
+g\chi_p\Omega L_0\nonumber
\\
&= \omega^2-\Omega^2 +\frac{i\Omega}{\tau_p}
+g\chi_p\Omega^2|A|^2
+g\chi_p\Omega(\Omega+\Omega_s)|B|^2\nonumber\\
&= \Omega_0^2 + \Omega^{\prime\: 2} |B|^2 \quad \text{where} \quad \Omega_0^2= \omega^2+\Omega^2 +i \Omega \Gamma_0 \quad \text{and} \quad \Omega^{\prime\: 2}=i \Omega \Gamma^{\prime}.
\end{align}

Plugging into \eqref{eq:dif_eq_for_B} gives:
\begin{align}
    \ddot{B} +\Gamma_0 \dot{B} +\Gamma^{\prime}|B|^2 \dot{B} + (\Omega_0^2 - \mathcal{S}_0) B + \Omega^{\prime\: 2}|B|^2 B =0.
\end{align}

Now, using the definition in \eqref{eq_Gamma_full} and \eqref{eq_oMEGa_full} and plugging into \eqref{eq:master}:

\begin{align}
&-\Omega_s^2
+i\Omega_s\left(\frac{1}{\tau_p}+i2\Omega
-i g\chi_p\Omega|A|^2
-i g\chi_p(\Omega+\Omega_s)|B|^2
\right)
+\frac{i\Omega}{\tau_p} +
\omega^2-\Omega^2 +g\chi_p\Omega^2|A|^2 \nonumber
\\
&+g\chi_p\Omega(\Omega+\Omega_s)|B|^2 =
\frac{-g\chi_p}{1+i\Omega_s\tau_s}
\left(\frac{2\Omega+\Omega_s}{2}\right)^2 |A|^2.
\end{align}

Collecting the $|B|^2$ terms:
\begin{align}
\label{eq:La_necesito}
g\chi_p(\Omega+\Omega_s)^2 |B|^2
&=
\frac{-g\chi_p}{1+i\Omega_s\tau_s}
\left(\frac{2\Omega+\Omega_s}{2}\right)^2 |A|^2
+\Omega_s^2-\frac{i\Omega_s}{\tau_p}+2\Omega\Omega_s -g\chi_p\Omega\Omega_s|A|^2 \nonumber
\\
&\quad
-\omega^2+\Omega^2
-g\chi_p\Omega^2|A|^2
-\frac{i\Omega}{\tau_p}.
\end{align}

Now, given that:
\begin{align}
\frac{1}{1+i\Omega_s\tau_s}
=
\frac{1}{1+\Omega_s^2\tau_s^2}
-i\,\frac{\Omega_s\tau_s}{1+\Omega_s^2\tau_s^2},
\end{align}
we can express Eq.~\ref{eq:La_necesito} as follows:
\begin{align}
    |B|^2 &= \alpha + i \beta, \quad \text{where }\alpha \text{ and } \beta \text{ are real numbers.}
\end{align}

Since $|B|^2$ is the square of a norm, it must be real and greater than or equal to zero. Therefore, we require that $\beta=0$. Then:
\begin{align}
    -\frac{\Omega_s}{\tau_p}-\frac{\Omega}{\tau_p} + \left(\frac{2\Omega+\Omega_s}{2}\right)^2 |A|^2 \frac{g \chi_p \Omega_s\tau_s}{1+\Omega_s^2\tau_s^2}=0.
\end{align}

Solving this equation for $\Omega_s$ provides another criterion to determine its value, similar to Eq.~\ref{eq:just_omega_s}. 
Using this condition, we obtain:
\begin{align}
\label{eq:B2_usefull}
|B|^2
&= \alpha \nonumber\\
&=\frac{1}{g\chi_p(\Omega+\Omega_s)^2}
\Bigg(
\Omega_s^2+2\Omega\Omega_s
-g\chi_p\Omega\Omega_s|A|^2
-\omega^2+\Omega^2
-g\chi_p\Omega^2|A|^2 \nonumber
\\
&\hspace{0.2cm}-\left(\frac{2\Omega+\Omega_s}{2}\right)^2 |A|^2\,g\chi_p\,
\frac{1}{1+\Omega_s^2\tau_s^2}
\Bigg).
\end{align}

This result reproduces the form of the angular momentum amplitude obtained numerically (see Fig. 3.(a) in the main text) near the transition point.

\begin{figure}[h!] 
\centering \includegraphics[width=6cm]{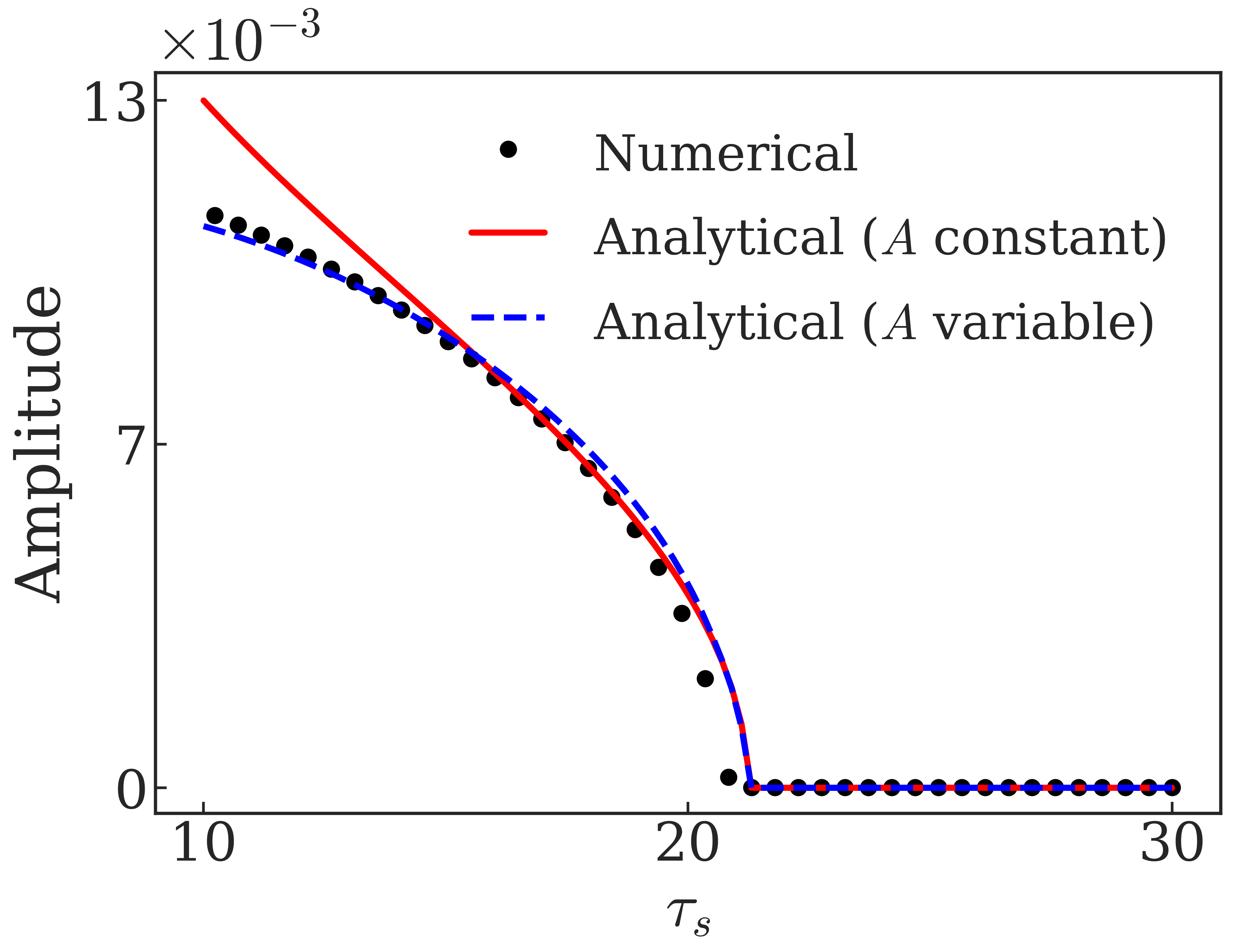}  \caption{
Comparison between the numerical angular momentum amplitude and analytical predictions as a function of the spin relaxation time $\tau_s$. The analytical result assuming constant $A$ captures the numerical behavior only near the transition, while allowing $A$ to vary yields quantitative agreement over a wider range of $\tau_s$. The parameters used are $\Omega = 1.16\omega$, $F = 0.08\omega\sqrt{\hbar \omega}$, $\chi_p = 4.0/\hbar$, $g = \chi_p(10/1.16^2)\hbar\omega$, and $\omega\tau_p = 10$.
}
\label{Comparation_amplitude} 
\end{figure}

A direct comparison between numerical and analytical results is shown in Fig.\ref{Comparation_amplitude}, where the numerical data is contrasted with analytical expressions obtained for different assumptions for the parameter $A$.

Here, we work in a regime where $|B_0|^2$ is not sufficiently small to be neglected, so one might be tempted to expect that the same functional form observed in the numerical results should be recovered for all values of $\tau_s$, and not only near the transition point. However, our analysis assumes that the parameter $A$ is constant with respect to $\tau_s$. This assumption is valid in the constant angular momentum regime, as shown in the subsection \ref{subsec:trial_solution_cp}(where $A$  is denoted by $|\bar{Q}|$ in Eq.\eqref{eq:Qbar_solution}) but it breaks down when $\tau_s$ is decreased beyond the transition point. 

This behavior is clearly illustrated in Fig.\ref{Comparation_amplitude}, where the analytical expression with constant $A$ is accurate only in the vicinity of the transition. Allowing $A$ to vary with $\tau_s$ give us better approximation in a  broader range. Those $\tau_s$-dependent A values were obtained from the numerical trajectories by performing Fourier transforms for each value of $\tau_s$. 

We rewrite Eq.~\eqref{eq:B2_usefull} as $B_0^2=f(A,\Omega)$, where $B_0\equiv |B|\ge 0$.
This allows us to introduce an effective (Landau-like) pseudo-potential $U(B_0)$ defined by
\begin{align}
\frac{\partial U}{\partial B_0} &= B_0\left(B_0^2-f\right),\nonumber\\
U(B_0) &= \frac{1}{4}B_0^4-\frac{1}{2}f\,B_0^2.
\end{align}
The extrema satisfy $B_0(B_0^2-f)=0$, so the nontrivial solution $B_0=\sqrt{f}$ exists only for $f>0$, therefore the phase boundary is given by $f(A(F,\Omega),\Omega)=0$.
We compute $A$ from Eq.~\eqref{eq:Qbar_solution} and obtain $\Omega_s$ from Eq.~\eqref{eq:just_omega_s}, which together determine $f$ over a grid of $(F,\Omega)$ and yield the boundary shown in Fig.~4 of the main text.
The analytical criterion matches the numerically extracted boundary over most of the parameter space, with noticeable deviations only in the corner of simultaneously small $F$ and $\Omega$, where the oscillation amplitude is very weak and the extraction of $\Omega_s$ becomes less reliable.

\bibliographystyle{unsrt}
\bibliography{ref,ref-driven-spin-phonon}

\end{document}